\documentclass[fleqn,usenatbib]{mnras}

\usepackage{newtxtext,newtxmath}

\usepackage[T1]{fontenc}
\usepackage{ae,aecompl}

\usepackage{graphicx}	
\usepackage{amsmath}	
\usepackage{amssymb}	

\newcommand{\psrb}{PSR\,B1259$-$63}

\title[PSR B1259-63 astrometry]{The geometric distance and binary orbit of PSR B1259-63}

\author[J. C. A. Miller-Jones et al.]{
J. C. A. Miller-Jones,$^{1}$\thanks{E-mail: james.miller-jones@curtin.edu.au (JCAM-J)}
A. T. Deller,$^{2}$
R. M. Shannon,$^{1,2,3,4}$
R. Dodson,$^{5}$
J. Mold\'on,$^{6}$
\newauthor
M. Rib\'o,$^{7}$\thanks{Serra H\'unter Fellow.} 
G. Dubus,$^{8}$
S. Johnston,$^{3}$
J. M. Paredes,$^{7}$
S. M. Ransom,$^{9}$
\newauthor
and J. A. Tomsick$^{10}$
\\
$^{1}$International Centre for Radio Astronomy Research -- Curtin University, GPO Box U1987, Perth, WA 6845, Australia\\
$^{2}$Centre for Astrophysics and Supercomputing, Swinburne University of Technology, John St, Hawthorn, VIC 3122, Australia\\
$^{3}$CSIRO Astronomy and Space Science, Australia Telescope National Facility, PO Box 76, Epping, NSW 1710, Australia\\
$^{4}$ARC Centre of Excellence for Gravitational Wave Discovery (OzGrav)\\
$^{5}$International Centre for Radio Astronomy Research, The University of Western Australia, 35 Stirling Highway, WA 6009, Australia\\
$^{6}$Jodrell Bank Centre for Astrophysics, Alan Turing Building, The University of Manchester, Oxford Road, Manchester, M13 9PL, UK\\
$^{7}$Departament de F\'{\i}sica Qu\`antica i Astrof\'{\i}sica,
Institut de Ci\`encies del Cosmos (ICCUB), Universitat de Barcelona,
IEEC-UB, \\Mart\'{\i} i Franqu\`es 1, E08028 Barcelona, Spain\\
$^{8}$Universit\'e Grenoble Alpes, CNRS, IPAG, 38000 Grenoble, France\\
$^{9}$National Radio Astronomy Observatory, 520 Edgemont Road, Charlottesville, Virginia 22903-2475, USA\\
$^{10}$Space Sciences Laboratory, 7 Gauss Way, University of California, Berkeley, CA 94720-7450, USA\\
}

\date{Accepted 2018 July 2. Received 2018 June 29; in original form 2018 April 23}

\pubyear{2018}

\begin{document}
\label{firstpage}
\pagerange{\pageref{firstpage}--\pageref{lastpage}}
\maketitle

\begin{abstract}
The pulsar/massive star binary system \psrb/LS\,2883 is one of the best-studied gamma-ray binaries, a class of systems whose bright gamma-ray flaring can provide important insights into high-energy physics. Using the Australian Long Baseline Array we have conducted very long baseline interferometric observations of \psrb\ over 4.4 years, fully sampling the 3.4-year orbital period. From our measured parallax of $0.38\pm0.05$\,mas we use a Bayesian approach to infer a distance of $2.6^{+0.4}_{-0.3}$\,kpc.  We find that the binary orbit is viewed at an angle of $154\pm3$\degr\ to the line of sight, implying that the pulsar moves clockwise around its orbit as viewed on the sky.  Taking our findings together with previous results from pulsar timing observations, all seven orbital elements for the system are now fully determined.  We use our measurement of the inclination angle to constrain the mass of the stellar companion to lie in the range 15--31$M_{\odot}$.  Our measured distance and proper motion are consistent with the system having originated in the Cen OB1 association and receiving a modest natal kick, causing it to have moved $\sim 8$\,pc from its birthplace over the past $\sim3\times10^5$ years.  The orientation of the orbit on the plane of the sky matches the direction of motion of the X-ray synchrotron-emitting knot observed by the {\em Chandra X-ray Observatory} to be moving away from the system.
\end{abstract}

\begin{keywords}
astrometry -- parallaxes -- proper motions -- pulsars: individual: \psrb\ -- radio continuum:stars -- gamma-rays: stars
\end{keywords}



\section{Introduction}

Gamma-ray binaries are systems comprised of a massive star in orbit with a compact object, whose broadband, non-thermal energy output peaks in the MeV--GeV band, with the emission typically being modulated on the orbital timescale.  Seven such binaries with confirmed high-energy gamma-ray emission are currently known \citep[see, e.g.,][]{Dubus17,Veritas17}, but only in two systems has the nature of the compact object been confirmed, via the detection of radio pulsations from the neutron star \citep{Johnston92,Lyne15}. The most well-studied of these two systems is \psrb/LS\,2883, comprising the radio pulsar \psrb\ in a wide, eccentric orbit \citep*[$P_{\rm orb}=1236.9$\,d, $e=0.87$;][]{Shannon14} with the rapidly-rotating, massive, late Oe-type companion star, LS\,2883 \citep{Negueruela11}.  While other gamma-ray binaries (LS I +61$^{\circ}$303 and HESS J0632+057) 
have similar stellar companions showing the Be phenomenon \citep*{Rivinius13}, their tighter orbits mean that even if the compact objects in those systems were otherwise observable radio pulsars, the enshrouding material would likely render the detection of radio pulsations impossible.  This makes \psrb/LS\,2883 a key system for understanding gamma-ray binaries, a class of system that can provide important insights into pulsar winds and binary evolution theory \citep[e.g.,][]{Dubus13}.

High-energy GeV and TeV flaring is detected from the \psrb/LS\,2883 system every orbit around periastron passage \citep[e.g.][]{Aharonian05,Aharonian09,Tam11} as the pulsar passes through \citep[and possibly disrupts; see, e.g.][]{Chernyakova15} the circumstellar disk of its O9.5Ve companion star.  The gamma-ray production mechanism has been suggested to be inverse Compton upscattering of stellar photons by relativistic electrons, either in the shocked region where the pulsar wind interacts with the stellar wind from the massive companion \citep*{Tavani94,Kirk99}, or in the unshocked pulsar wind \citep{Petri11,Khangulyan11}.  Such inverse Compton emission depends on the scattering angle between the electrons and the stellar photons, and would thus be subject to relativistic aberration and boosting \citep*{Dubus10}.  Alternatively, it has been proposed that the GeV flares could be attributed to Doppler-boosted synchrotron emission in the bow-shock tail formed as the shocked pulsar wind is collimated by the stellar wind \citep*{Dubus10,Kong12}.  Regardless of the exact mechanism, the observed emission will depend on the orbital parameters of the system, in particular the inclination to the line of sight.

When characterising the \psrb/LS\,2883 system, there are three key angles to consider; the inclinations of the binary orbit, the pulsar rotation axis, and the Oe companion star disk to the line of sight.  Due to the potential natal kick experienced during the supernova in which the neutron star was formed \citep[e.g.][]{Lyne94}, these angles need not all be aligned.  The orbital inclination has been estimated via several methods.  From the measured mass function, \citet{Johnston94} assumed likely component masses to derive an inclination angle of $\sim35$\degr.  \citet{Manchester95a} modelled the pulse profiles and polarization parameters to determine that the pulsar rotation axis was inclined at $46\pm6$\degr\ to the line of sight.  Instead, \citet{Negueruela11} used optical observations of the massive companion star to infer that the stellar rotation axis \citep*[which is believed to be inclined at an angle of $\sim10$\degr\ to the orbital plane;][]{Melatos95} was inclined at 33\degr\ to the line of sight.  Using the inferred stellar mass and the measured mass function of the system, they deduced an orbital inclination angle of $i=25^{+6}_{-5}$\degr. Finally, \citet{Shannon14} found the orbital inclination to the line of sight to be $i=154\pm4$\degr, and the inclination of the companion star spin axis to be $147\pm3$\degr, with a $35\pm7$\degr\ misalignment between the spin and orbital angular momenta (noting the three-dimensional geometry of the system). Since the sense of rotation is not known from any existing observation, inclination angles of $i$ and $180^{\circ}-i$ are currently degenerate, such that the estimates of $25^{+6}_{-5}$\degr\ and $154\pm4$\degr\ derived by \citet{Negueruela11} and \citet{Shannon14}, respectively, are consistent within uncertainties.  Henceforth, we adopt the standard convention that inclinations of $0\degr <i<90\degr$ denote counterclockwise orbits, and $90\degr<i<180\degr$ denote clockwise orbits.
An independent measurement of the orbital inclination would allow us to deproject the semi-major axis of the pulsar orbit and hence improve our estimates of the component masses via Kepler's Third Law.  In addition, together with the distance it would help discriminate between the different models for the gamma-ray emission mechanism.

\citet{Johnston92} originally assumed a distance of 2.3\,kpc to \psrb, based on its dispersion measure and the model of the Galactic electron density derived by \citet*{Lyne85}.  The revised electron density model of \citet*{Taylor93} suggested a higher distance of 4.6\,kpc, although this was ruled out by \citet{Johnston94} based on the implausible parameters this distance would imply for the nature of the optical companion.  Instead, \citet{Johnston94} assumed a location in the Sagittarius arm, giving a distance of 1.5\,kpc \citep{Georgelin76}.  The most recent and widely adopted distance constraint comes from spectral and photometric studies indicating that the system lies at the same distance as the Cen OB1 association, implying a distance of $2.3\pm0.4$\,kpc \citep{Negueruela11}.

The only model-independent method of distance determination is via geometric parallax.  This can be accomplished either via optical observations of the companion star \citep[e.g.\ with {\it Gaia};][]{Gaia16}, or by Very Long Baseline Interferometry (VLBI) using radio observations of the pulsar.  While VLBI parallax measurements are routine for systems such as young stars \citep[e.g.][]{Loinard05}, X-ray binaries \citep[e.g.][]{Miller-Jones09} and pulsars \citep[e.g.][]{Deller09}, in the case of \psrb\ the proximity and long orbital period of the system imply that any parallax measurement will need to be disentangled from the orbital signature \citep[see, e.g.,][]{Tomsick10}. While orbital motion has been detected in a few closer pulsar systems \citep[e.g.][]{Deller13,Deller16}, most pulsar orbits are too small to be accurately constrained by VLBI.

At a distance of 2.3\,kpc, the projected semi-major axis of the pulsar orbit in the \psrb/LS\,2883 system \citep[1296.27448(14) light seconds;][]{Shannon14} implies that for an inclination angle of 24.7\degr, a companion mass of $30M_{\odot}$ \citep{Negueruela11} and an assumed neutron star mass of $1.4M_{\odot}$, we would expect the semi-major axis of the pulsar orbit to subtend 2.7 milliarcseconds (mas) on the sky, and that of the companion 0.13\,mas, as compared to a predicted parallax of 0.43\,mas.  Therefore, any geometric distance determined from a simple five-parameter fit (for position, proper motion and parallax) will be significantly in error, and a full solution will require observations sampling not only the parallax ellipse, but also the full 3.4-year orbit.

A model-independent geometric distance would help constrain the optical luminosity of the companion star, and hence the availability of seed photons and therefore the expected gamma-ray luminosity from inverse Compton upscattering.  During the large GeV flare observed 30\,d post-periastron in 2010 \citep{Abdo11,Tam11}, the observed gamma-ray flux implied a luminosity of $8\times10^{35} (d/2.3 {\rm\, kpc})^2$\,erg\,s$^{-1}$, very close to the available pulsar spin-down luminosity of $8.2\times10^{35}$\,erg\,s$^{-1}$ \citep{Manchester95} and thereby posing a severe constraint on gamma-ray emission models \citep{Khangulyan12,Dubus13a}.  While similar GeV luminosities were observed in the subsequent periastron passage of 2014, the peak GeV flux reported during the 2017 periastron passage was a factor of two higher \citep{Johnson17}, favouring models involving relativistic boosting.

In conjunction with a proper motion measurement, a geometric distance would also enable confirmation of the proposed birthplace of the system in the Cen OB1 association \citep{Shannon14}, and provide constraints on any natal kick the system might have received during the supernova in which the neutron star was formed.  The proper motion of the system was initially constrained by pulsar timing data \citep{Shannon14} to be $-6.6\pm1.8$\,mas\,yr$^{-1}$ in Right Ascension (R.A.) and $-4.4\pm1.4$\,mas\,yr$^{-1}$ in Declination (Dec.).  This was subsequently revised by {\it Gaia} Data Release 1 (DR1) to $-6.5\pm0.6$\,mas\,yr$^{-1}$ in R.A.\ and $0.1\pm0.7$\,mas\,yr$^{-1}$ in Dec.\ \citep{Lindegren16}.  The cause of this discrepancy in the Dec.\ component is unexplained, but could potentially be due to incomplete treatment of stochastic parameters in pulsar timing data, such as rotational irregularities or dispersion measure variations \citep[e.g.][]{Lentati14}.

{\it Gaia} Data Release 2 (DR2) further refined the estimated proper motion of the optical component of the system, LS\,2883, and provided an initial estimate of the parallax. Using a standard five-parameter astrometric fit applied to data taken between 2014 July 25 and 2016 May 23, the measured proper motions were found to be $-6.99\pm0.04$\,mas\,yr$^{-1}$ in R.A.\ and $-0.42\pm0.04$\,mas\,yr$^{-1}$ in Dec.\ \citep{Lindegren18}.  The fitted parallax was $0.42\pm0.03$\,mas, which would correspond to an inversion distance of $2.4\pm0.2$\,kpc \citep[although see][for a description of how a Bayesian approach should instead be used to infer distances from {\it Gaia} parallax measurements]{Luri18}.  However, as described above we expect the semi-major axis of the orbit of LS\,2883 to subtend of order 0.1\,mas on the sky, such that the unmodelled orbital motion could significantly affect the fitted astrometric parameters.  This expectation is confirmed by the poor goodness-of-fit statistic of the {\it Gaia} DR2 astrometric solution, whose high value of 7.9 (as compared to the expected normal distribution of zero mean and unit standard deviation) indicates an unsatisfactory fit to the data. At this time, we therefore consider the {\it Gaia} results primarily as a cross-check on the new VLBI results that we present in this paper, and we provide a brief comparison in Section \ref{sec:discussion}.

The radio emission of the \psrb/LS\,2883 system consists of two components, one from the pulsar and the other from a more extended synchrotron-emitting nebula. The pulsed emission (which provides a point-like astrometric probe) has a double-peaked profile, with a flux density of 0.3--1.9 mJy at 8.4 GHz \citep{Johnston05}. The pulsations disappear for $\sim 20$ days either side of periastron passage, likely due to the eclipsing of the pulsations by the equatorial circumstellar disk surrounding the massive star \citep{Melatos95}. Close to periastron passage, transient, unpulsed synchrotron emission is also seen, likely due to the strong interaction between the stellar wind and the pulsar wind. The shocked material expands adiabatically, producing a nebula extending 50--100\,mas away from the stellar companion \citep{Moldon11,Chernyakova14}. This diffuse emission has been detected up to 100 days after periastron.  To avoid systematic astrometric offsets due to the contribution of this diffuse unpulsed emission, VLBI astrometry should use the pulsed emission alone.

We note that the orbit of \psrb\ shows strong evidence for precession of the orbital plane due to classical spin-orbit coupling \citep{Wex98}. However, the rate of periastron advance is $(7.81\pm0.03)\times10^{-5}$\degr\,yr$^{-1}$ \citep{Shannon14}, which is not detectable in astrometric VLBI observations.

In this paper, we present a 4.4-year campaign of astrometric VLBI observations of \psrb, using the Australian Long Baseline Array (LBA).  We detail our observations and data reduction in Section~\ref{sec:obs}, and present our results in Section~\ref{sec:results}.  We discuss their implications for the system as a whole in Section~\ref{sec:discussion}, and draw our conclusions in Section~\ref{sec:conclusions}.

\section{Observations and data reduction}
\label{sec:obs}

Using the LBA, we made eleven observations of \psrb\ over a 4.4-year period, sampling more than one full 1236-day binary orbit, as detailed in Table~\ref{tab:obslog}.  The trajectory of \psrb\ on the sky is made up of the proper motion, parallax and orbital signatures, all of which we aimed to sample during our astrometric campaign.

The LBA typically observes in three to four week-long sessions each year.  However, to sample the maxima and minima of the parallax ellipse and the epochs closest to periastron passage, it was necessary to schedule a few of our epochs in out-of-session observing runs with a reduced number of stations.  Operational constraints at the Parkes 64-m telescope (limited receiver changes) and Tidbinbilla (availability in between Deep Space Network observations) meant that these telescopes were not available for all observations, even in-session.  As a consequence, our array varied significantly between epochs (Table~\ref{tab:obslog}), leading to differences in {\it uv}-coverage and resolution.  The frequency setup (bandwidth and central frequency) also varied between epochs, according to what mode could be accommodated at the time of each observation, as detailed in Table~\ref{tab:obslog}.  Our central frequency ranged from 8.425--8.477\,GHz, and we observed in dual polarisation in all cases.

We observed in phase referencing mode, using the nearby extragalactic source J1256$-$6449 \citep[1.23\degr\ from \psrb;][]{Petrov11} as our phase reference calibrator.  We fixed its assumed position to be the same in each epoch, at R.A.\ (J2000) = 12$^{\rm h}$56$^{\rm m}$03\fs 4032, Dec.\ (J2000) = $-$64$^{\rm d}$49$^{\prime}$14\farcs 814\footnote{Position taken from \url{http://astrogeo.org/}, using the rfc\_2010c solution.}.  All target positions were derived relative to this assumed calibrator position.  The phase referencing cycle time was 5\,min, with 3.5\,min on the target and 1.5\,min on the calibrator in each cycle.

\begin{figure}
	\includegraphics[width=\columnwidth]{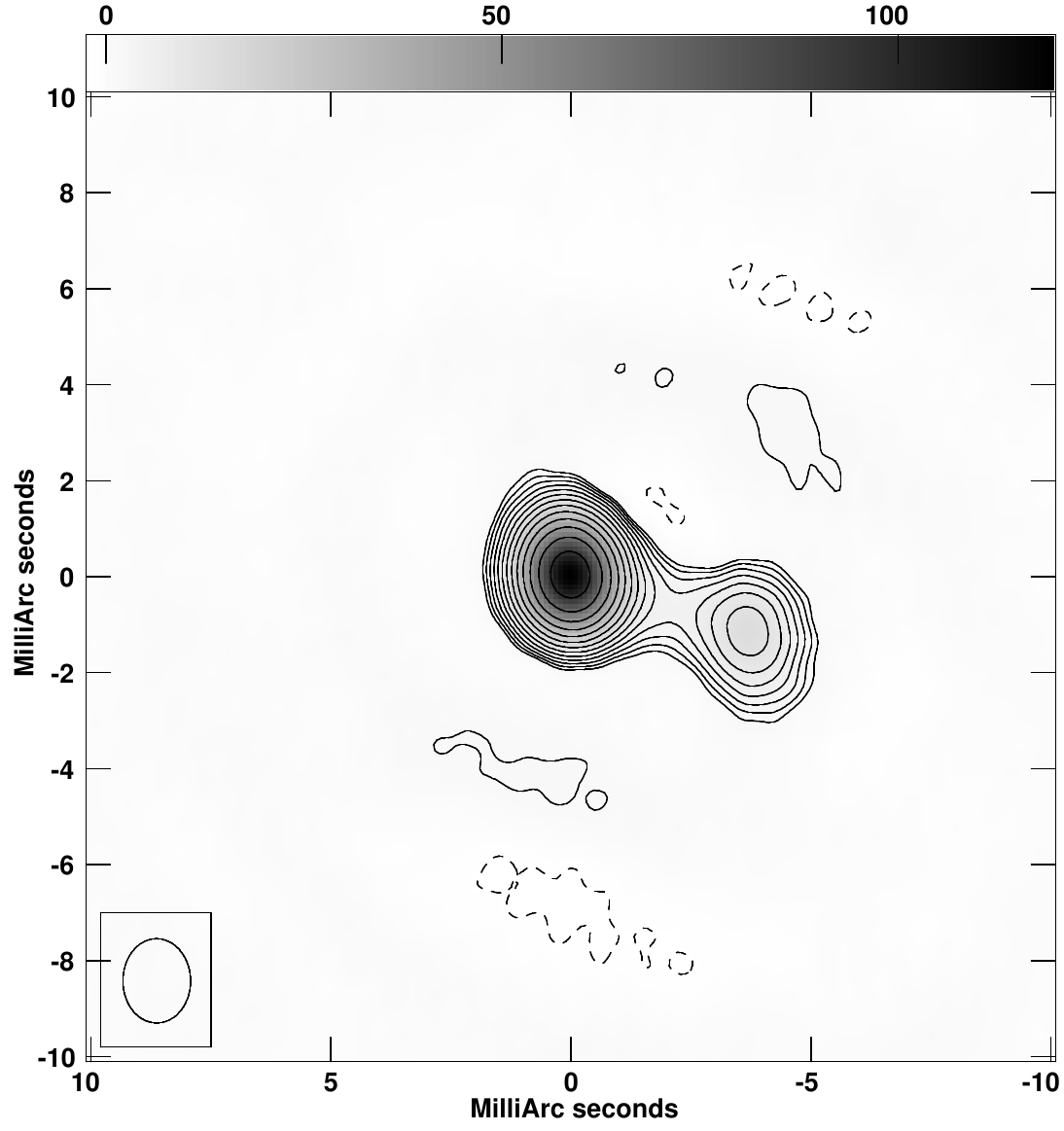}
    \caption{Stacked image of the phase reference calibrator, J1256$-$6449. Contours are at levels of $\sqrt{2}^n$ times the $3\sigma$ noise level of 1.5\,mJy\,beam$^{-1}$, where $n=0,1,2,...$ The peak brightness of the compact core component is 119\,mJy\,beam$^{-1}$.}
    \label{fig:j1256}
\end{figure}

\begin{table*}
	\footnotesize
	\centering
	\caption{Observing log for our LBA observations of \psrb.  $T-T_0$ denotes time since the most recent periastron passage, in days.  The bandwidth column details the number of baseband channels, and the bandwidth of each of those channels (in MHz).  Stations include a single antenna from the Australian Square Kilometre Array Pathfinder (ASKAP; Ak), the phased-up Australia Telescope Compact Array (ATCA; At), Ceduna (Cd), Hartebeesthoek 25m (Hh), Hobart (Ho), Hartebeesthoek 15m (Ht), Katherine (Ke), Mopra (Mp), Parkes (Pa), the Tidbinbilla antennas (either the 70m, DSS43, or one of the 34m telescopes DSS34, DSS35, DSS45), Warkworth 12m (Ww), Warkworth 30m (Wa), and Yarragadee (Yg).}
	\label{tab:obslog}
	\begin{tabular}{lccccl} 
		\hline
		Epoch & Date & MJD & $T-T_0$ & Bandwidth & Array\\
        & (UT) & & (d) & (MHz) & \\
		\hline
		A & 2013 Mar 17, 09:00--21:00 & $56368.62\pm0.25$ & \phantom{1}823.9 & $8\times8\phantom{1}$ & Ak, At, Cd, Hh, Ho, Ht, Ke, Pa, DSS34, DSS45, Ww\\
		B & 2013 Aug 15, 22:00--15:00 & $56520.21\pm 0.24$ & \phantom{1}975.5 & $8\times8\phantom{1}$ & Ak, At, Cd, Hh, Ho, Ke, Mp, Pa, DSS45, Ww, Yg\\
		C & 2014 Apr 03, 08:00--20:00 & $56750.58\pm0.25$ & 1205.9 &  $2\times16$ & At, Cd, Ho, Ke, Mp, DSS43, Ww, Yg\\
		D & 2014 Jun 03, 03:44--15:00 & $56811.40\pm0.22$ & \phantom{10}30.0 & $4\times8\phantom{1}$ & Ak, At, Cd, Ho, Mp, Pa, Ww\\
		E & 2015 Mar 25, 07:44--20:00 & $57106.58\pm0.25$ & \phantom{1}325.2 & $8\times8\phantom{1}$ & Ak, At, Cd, Ho, Ke, Mp, Pa, DSS35, Yg\\
		F & 2015 Jul 15, 23:30--13:00 & $57219.29\pm0.25$ & \phantom{1}437.9 &$8\times8\phantom{1}$ & Ak, At, Cd, Ho, Ht, Mp, DSS35, DSS45\\
		G & 2015 Sep 27, 20:48-09:00 & $57293.13\pm0.25$ & \phantom{1}511.7 & $8\times16$ & Ak, At, Cd, Ho, Ht, Ke, Mp, Pa, Ww, Yg\\
		H  & 2016 Jan 21, 12:07--23:51 & $57408.75\pm0.24$ & \phantom{1}627.3 & $8\times16$ & Ak, At, Cd, Hh, Ho, Ke, Ww, DSS43\\
		I & 2016 Jun 26, 02:52--14:30 & $57565.36\pm0.24$ & \phantom{1}783.9 & $4\times16$ & Ak, At, Cd, Hh, Ho, Ke, Pa, Ww, Yg\\
		J & 2017 Jan 20, 13:52--01:00  & $57773.81\pm0.22$ & \phantom{1}992.4 & $4\times16$ & At, Cd, Hh, Ho, Ke, Mp, DSS35, Ww, Yg\\
		K & 2017 Aug 11, 22:52--11:00 & $57977.21\pm0.24$ & 1195.8 & $4\times16$ & At, Cd, Hh, Ho, Ke, Mp, Pa, DSS43, Wa, Yg\\
		\hline
	\end{tabular}
\end{table*}

Data were correlated using the DiFX software correlator \citep{Deller07}, and were reduced according to standard procedures within the Astronomical Image Processing System \citep[AIPS, version 31DEC17;][]{Greisen03}.  We first corrected the visibility amplitudes for errors in sampler threshold levels using the autocorrelation spectra, and then calibrated the visibility amplitudes using either measured system temperatures (where available), or nominal values for the antennas with no recorded system temperature information.  For the first nine epochs, station velocities for the antennas at Katherine, Warkworth and Yarragadee were not available at the time of correlation, leading to station position errors of up to 93\,cm.  Given the wavelength of 3.6\,cm and the calibrator throw of 1.23\degr, this is equivalent to a change of phase of up to $\Delta \phi = \frac{\Delta r}{\lambda}\sin\theta = 33$\degr, where $\Delta r$ is the position shift, $\lambda$ is the observing wavelength, and $\theta$ is the calibrator throw.  We therefore corrected the antenna positions in our raw data using the best available velocities before proceeding with further calibration.

Next, we corrected for ionospheric delay using a map of the ionospheric electron content.  We solved for the instrumental delays and rates using bright fringe finder sources (B0208$-$512, B0537$-$441, B0637$-$752, B1424$-$418, or B1610$-$771).  We flagged any radio frequency interference, and discarded all data taken at elevations of $<20$\degr, where the longer path through the atmosphere and ionosphere leads to unacceptably large calibration errors.  We then performed bandpass and preliminary amplitude calibration using a bright fringe finder source.  Next, we conducted global fringe fitting on the phase reference source to solve for phases, delays and rates.  We assumed a point source model, and maximized the signal-to-noise ratio on the small long-baseline stations by summing the data across all frequencies and polarizations.

We generated calibrated data on the phase reference calibrator from each epoch, and stacked the common frequency subbands (spanning 8.409--8.441\,GHz) from these data sets together to enable hybrid mapping of J1256$-$6449.  We aligned the amplitude scales via one round of amplitude-and-phase self-calibration with a one-day solution interval.  Following two rounds of phase-only self-calibration on timescales of 2 and 0.5\,min, we applied one further round of amplitude-and-phase self-calibration with a 30-min solution interval before generating the final model image shown in Fig.~\ref{fig:j1256}.  This global model was then used to rerun the fringe-fitting process on J1256$-$6449 for each epoch.  This allowed us to take account of the source structure, with the use of a single global model image for all data sets preventing the slight differences in the per-epoch models due to the changing array configuration from generating spurious astrometric shifts.  Finally,  for each epoch we used the same global calibrator model to conduct a final round of amplitude-and-phase self-calibration on a 30-minute timescale, applying only these short-timescale amplitude solutions to the data.

\subsection{Pulsar binning}
\label{sec:binning}
The pulse profile of \psrb\ is known to be double-peaked \citep{Johnston92}.  We therefore used the DiFX software correlator \citep{Deller07} to provide both an unbinned data set, and four pulse phase-resolved bins to represent the full pulse profile; two on-pulse bins comprising a total of 20 per cent of the pulse period, and two off-pulse bins making up the rest.  Unfortunately, an erroneous correlator setup for Epoch B meant that no pulsar binned data were available for that observation for the long-baseline stations at ASKAP, Katherine, Warkworth or Yarragadee.  This reduced the available baseline length, and hence the astrometric accuracy of this epoch.

Having calibrated the continuum data for each epoch as detailed above, we applied the derived solution tables to the corresponding four sets of binned data.  We weighted the individual bins by the inverse square root of the bin duration, and imaged each bin separately.  With the exception of a single off-pulse bin from epoch D (taken 30\,d after periastron; see Section~\ref{sec:unpulsed}), we saw no evidence for emission in the off-pulse bins.  We can therefore be confident that all the radio emission arises from the pulsar, which should be a point-like astrometric source. To maximize the signal-to-noise ratio we combined the two on-pulse bins for each epoch (after reweighting to account for bin widths) when making our final images.

\subsection{Station selection and data weighting}
While we were able to correct the positions of the geodetic antennas Katherine, Warkworth and Yarragadee, the position of the ASKAP antenna remained relatively poorly determined (uncertainties of 0.2--0.4\,m), so we chose to discard ASKAP when making our final images.  We also discarded the Hartebeesthoek stations for all except the final three epochs (I, J, K), as those were the only three epochs in which the phase solutions at Hartebeesthoek could be reliably tracked between adjacent phase reference calibrator scans.

The most sensitive stations (Parkes, the Tidbinbilla 70-m telescope DSS43, and the phased ATCA) are all in the geographical centre of the array, and correspond to the shortest baselines.  Their high data weights translate to large synthesised beams and cause the astrometric positions to be disproportionately affected by any uncorrected systematic errors affecting those stations.  We therefore chose to use robust weighting for the initial imaging, with a robustness parameter of 0, to increase the weighting of the more distant stations (which tended to be least sensitive).  While this reduced the signal-to-noise ratio of the data, the pulsar was sufficiently bright in the binned data that the statistical errors remained similar to or lower than the systematic errors (see Section~\ref{sec:systematics}).

\section{Results and analysis}
\label{sec:results}

\begin{figure}
	\includegraphics[width=\columnwidth]{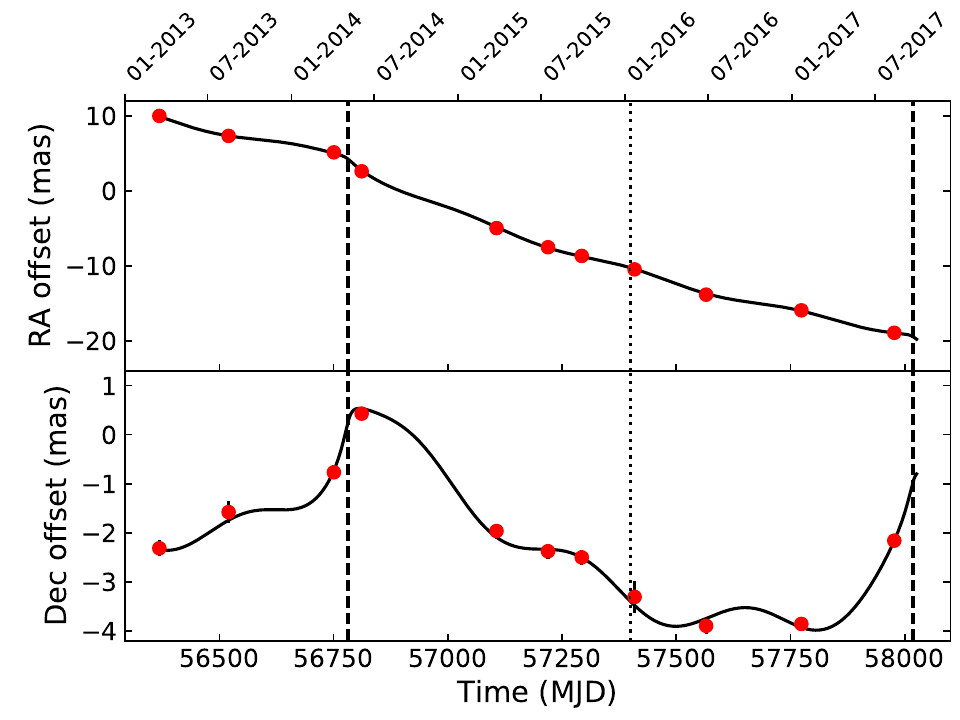}
    \caption{Motion of \psrb\ on the sky over time, in both R.A.\ (top) and Dec.\ (bottom).  The black line shows our best fitting astrometric solution.  The dashed and dotted vertical lines show the epochs of periastron and apastron, respectively. Our observations sample over one full orbital period, with three of the observations taken close to periastron.}
    \label{fig:sky_motion}
\end{figure}

The pulsed emission was significantly detected in all epochs, providing a point-like astrometric probe that allowed us to trace the motion of the system on the plane of the sky, as shown in Fig.~\ref{fig:sky_motion}.  While the proper motion dominates the trajectory in R.A., both the orbital signature and the annual parallax modulation can be seen in the Dec.\ trajectory.

\begin{table*}
	\centering
	\caption{Image weights, astrometric uncertainties, measured positions and flux densities for the stacked on-pulse LBA data for \psrb. Column (2) shows the weighting used for imaging (number denotes the robustness parameter, `S' shows that the square roots of the data weights were used to downweight highly sensitive baselines).  Columns (3) and (4) show the statistical uncertainty in R.A.\ ($\Delta\alpha_{\rm stat}$) and Dec.\ ($\Delta\delta_{\rm stat}$), in microarcseconds.  Column (5) shows the estimated systematic uncertainty, $\Delta\theta_{\rm sys}$ (assumed to be the same in R.A.\ and Dec.).  Columns (6) and (7) show the measured positions in R.A.\ and Dec., with the quoted total error from adding the statistical and systematic errors in quadrature. Column (8) shows the flux densities for the stacked data from the on-pulse bins from each epoch, with appropriate weighting for each bin as described in Section~\ref{sec:binning}. These are not the pulse-profile averaged flux densities, and are given primarily to show the signal-to-noise ratio of each observation.}
	\label{tab:measurements}
	\begin{tabular}{lccccccc}
		\hline
		Epoch & Image weights & $\Delta\alpha_{\rm stat}$ & $\Delta\delta_{\rm stat}$ & $\Delta \theta_{\rm sys}$ & Right Ascension & Declination & Flux density\\
		 & & ($\mu$as) & ($\mu$as) & ($\mu$as) & (J2000) & (J2000) & (mJy)\\
		\hline
A & \phantom{-}0,S &  \phantom{1}71 & \phantom{1}71 & 131 & 13$^{\rm h}$02$^{\rm m}$47\fs $63985\pm 0.00003$ & $-$63$^{\circ}$50$^{\prime}$08\farcs $6309\pm 0.0002$ & \phantom{1}$9.7\pm0.6$\\
B & \phantom{-}0\phantom{,S} & 166 & 159 & 129 & 13$^{\rm h}$02$^{\rm m}$47\fs $63944\pm 0.00004$ & $-$63$^{\circ}$50$^{\prime}$08\farcs $6302\pm 0.0002$ & $15.0\pm1.8$\\
C & -1\phantom{,S} &  \phantom{1}89 & \phantom{1}97 & 126 & 13$^{\rm h}$02$^{\rm m}$47\fs $63911\pm 0.00002$ & $-$63$^{\circ}$50$^{\prime}$08\farcs $6294\pm 0.0001$ & \phantom{1}$3.6\pm0.4$\\
D & -1\phantom{,S} &  \phantom{1}88 & 105 & \phantom{1}91 & 13$^{\rm h}$02$^{\rm m}$47\fs $63873\pm 0.00002$ & $-$63$^{\circ}$50$^{\prime}$08\farcs $6282\pm 0.0001$ & \phantom{1}$3.6\pm0.4$\\
E & \phantom{-}0\phantom{,S} &  \phantom{1}87 & 106 & 120 & 13$^{\rm h}$02$^{\rm m}$47\fs $63759\pm 0.00002$ & $-$63$^{\circ}$50$^{\prime}$08\farcs $6306\pm 0.0002$ & \phantom{1}$5.1\pm0.5$\\
F & -1\phantom{,S} &  \phantom{1}85 & \phantom{1}82 & 104 & 13$^{\rm h}$02$^{\rm m}$47\fs $63720\pm 0.00002$ & $-$63$^{\circ}$50$^{\prime}$08\farcs $6310\pm 0.0002$ & \phantom{1}$5.6\pm0.4$\\
G & \phantom{-}0,S &  \phantom{1}98 & \phantom{1}97 & 127 & 13$^{\rm h}$02$^{\rm m}$47\fs $63703\pm 0.00002$ & $-$63$^{\circ}$50$^{\prime}$08\farcs $6311\pm 0.0002$ & \phantom{1}$5.2\pm0.5$\\
H & \phantom{-}0\phantom{,S} & 139 & 305 & 126 & 13$^{\rm h}$02$^{\rm m}$47\fs $63675\pm 0.00003$ & $-$63$^{\circ}$50$^{\prime}$08\farcs $6319\pm 0.0003$ & \phantom{1}$5.2\pm1.2$\\
I & \phantom{-}0\phantom{,S} &  \phantom{1}91 & 123 & 122 & 13$^{\rm h}$02$^{\rm m}$47\fs $63624\pm 0.00002$ & $-$63$^{\circ}$50$^{\prime}$08\farcs $6325\pm 0.0002$ & \phantom{1}$1.6\pm0.2$\\
J & \phantom{-}0\phantom{,S} &  \phantom{1}74 & \phantom{1}99 & 118 & 13$^{\rm h}$02$^{\rm m}$47\fs $63593\pm 0.00002$ & $-$63$^{\circ}$50$^{\prime}$08\farcs $6324\pm 0.0001$ & \phantom{1}$5.5\pm0.6$\\
K & \phantom{-}0,S &  \phantom{1}42 & \phantom{1}60 & 119 & 13$^{\rm h}$02$^{\rm m}$47\fs $63548\pm 0.00001$ & $-$63$^{\circ}$50$^{\prime}$08\farcs $6307\pm 0.0001$ & \phantom{1}$5.1\pm0.4$\\
		\hline
	\end{tabular}
\end{table*}

\subsection{Systematic uncertainties}
\label{sec:systematics}

The systematic errors were calculated by taking into account all the atmospheric contributions; those of the static ionosphere, the static troposphere, the dynamic ionosphere and the dynamic troposphere. The latter two, having zero mean by definition, have little influence on the astrometric accuracy, although these set the time scale for successful phase connection. The static contributions `shift' the apparent source position for any antenna pair. Where the static contributions are coherent across the array (as for an atmospheric wedge) these contributions shift the source without noticeable loss of signal. 
This makes their existence and impact hard to assess from the data alone, so we resorted to a theoretical estimation of their impact.

We assumed typical values for the key parameters (e.g.\ 3\,cm of residual path length at zenith for the troposphere, and 6\,TEC units of residual total electron content (TEC) for the ionosphere), with the switching time and target/calibrator separation taken from the schedule. 
We used these values in the astrometric formulae of \citet{Asaki07} to assess the typical error terms on every baseline, taking into account the zenith angle of the source for each station. These errors, expressed as an effective path length (true path length for non-dispersive contributions such as the troposphere, frequency-dependent path length for the dispersive ionospheric contributions), were compared with the effective baseline length as determined from the restored beam size in the image.  As we used robust weighting this beam size was given not by the longest baseline but from a weighted sum of the various baseline contributions. The final systematic uncertainty for each epoch was then taken to be the expected error in the measurement divided by the effective baseline length.

Having derived the expected systematic uncertainties on the astrometric positions, we compared these values with the statistical uncertainties derived from our image-plane fitting.  In most cases, the error budget was dominated by the systematics, so we modified the robustness parameter and the intrinsic data weighting to reduce the restored beam size (and hence the calculated systematic uncertainty) at the expense of signal-to-noise ratio (and hence the measured statistical uncertainty).  While we would ideally have used the same weighting scheme across all epochs, the variation introduced by the presence or absence of the larger LBA telescopes meant that any given fixed approach would have given sub-optimal results in some cases.  Indeed, for epochs C, H and K it was necessary to remove the most sensitive baseline (phased ATCA--Tidbinbilla DSS43) entirely to prevent the measured position being dominated by any uncorrected systematics on that single highly-weighted baseline.  The minimum total uncertainty was achieved when the statistical and systematic contributions were as close to being equal as possible.  The final weighting schemes are reported in Table~\ref{tab:measurements}.  Having settled on the appropriate weighting for each epoch, we made the final images.  Following standard practise for VLBI astrometry, we used an image-plane fit to measure the final source positions and their statistical uncertainties, as fitting a model to the visibility data does not provide reliable uncertainties on the fitted parameters.  The statistical uncertainties were combined in quadrature with the systematic uncertainties to give the total uncertainties, which we report together with the fitted positions in Table~\ref{tab:measurements}.

\subsection{Fitting an astrometric model}
\label{sec:mcmc}

Five of the seven orbital elements (orbital period $P$, time of periastron $T_0$, eccentricity $e$, semi-major axis of the neutron star orbit $a_{\rm NS} \sin i$, and the argument of periastron, $\omega$) are already constrained to extremely high precision via pulsar timing \citep{Shannon14}.  We therefore needed to solve for the seven remaining astrometric parameters; reference position ($\alpha_0$, $\delta_0$), proper motion ($\mu_{\alpha}\cos\delta$, $\mu_{\delta}$), parallax $\pi$, longitude of the ascending node $\Omega$, and the inclination angle of the orbit to the line of sight $i$. 

\begin{figure*}
	\includegraphics[width=\textwidth]{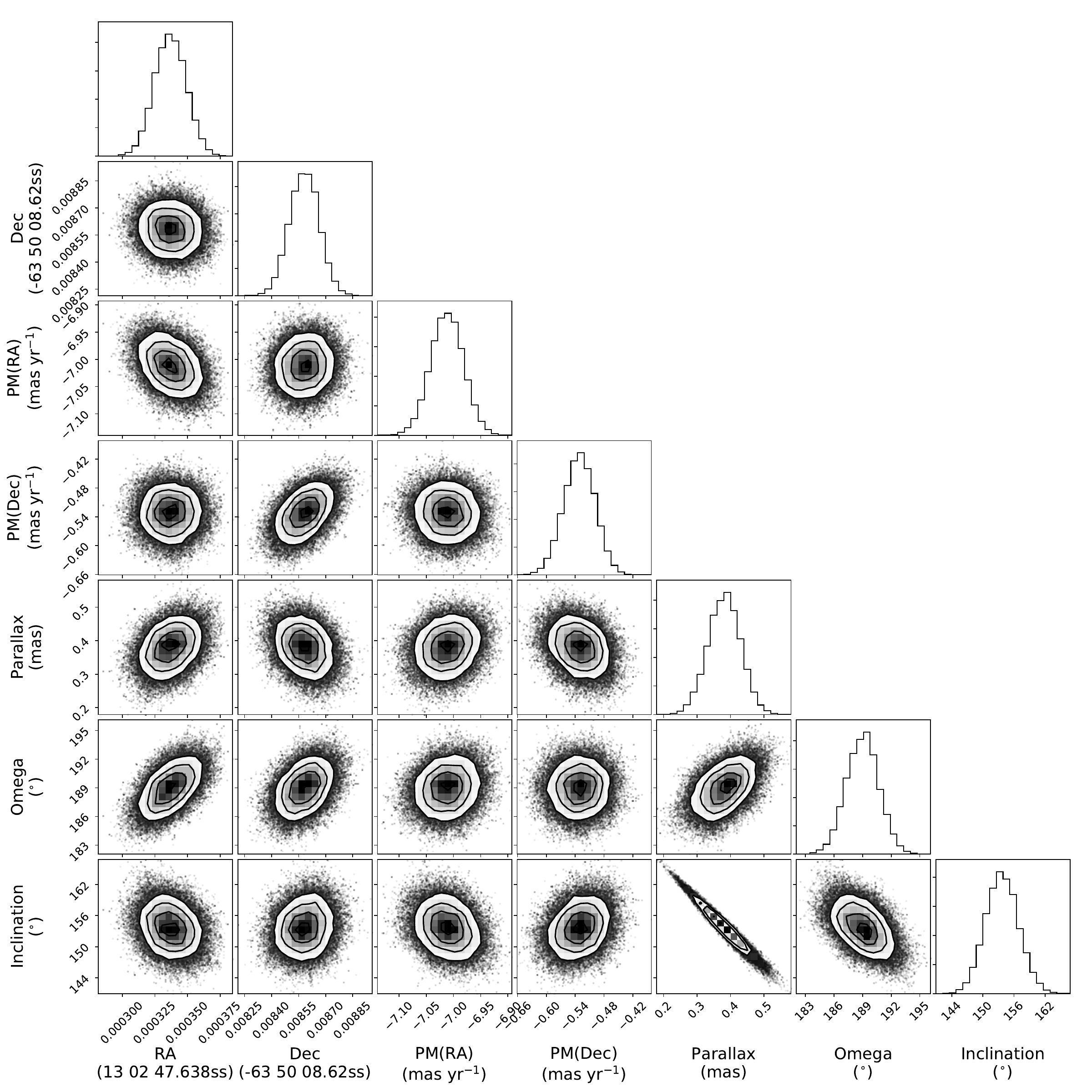}
    \caption{Results of the MCMC fit, showing our 130,000 samples projected in each one-dimensional (histograms) and two-dimensional (contour plots) representation of the overall seven-dimensional parameter space, revealing both the spread of results and their covariances.  We determine the position ($\alpha_0, \delta_0$), proper motion ($\mu_{\alpha}\cos\delta$, $\mu_{\delta}$), parallax ($\pi$), longitude of the ascending node ($\Omega$) and inclination angle ($i$) of \psrb, although as evident from the slope of the $\pi-i$ plot, there is a degeneracy between parallax and inclination angle.  Note that $\Omega$ is defined in the conventional notation of degrees counterclockwise from north through east.}
    \label{fig:triangle}
\end{figure*}

To fit for the seven unknown orbital and astrometric parameters, we used a Markov-Chain Monte Carlo (MCMC) algorithm, as implemented in the python package \textsc{emcee} \citep{Foreman-Mackey13}. We used the Naval Observatory Vector Astrometry Software \citep[NOVAS;][]{Barron11} with the DE421 solar system ephemeris to predict the parallax displacement of the pulsar, and the {\verb+binary_psr+} module of {\tt PRESTO} \citep{Ransom01} to calculate orbital reflex motion.  We used non-constraining uniform priors for all of the fitted parameters (with the inclination prior being uniform in $\cos i$ rather than $i$), and initialised the walkers using the best-fit values and uncertainties for the astrometric parameters obtained with a least-squares fit\footnote{ Our full fitting code is available at \url{https://github.com/adamdeller/astrometryfit}.}.  The results of the MCMC fitting are given in Table~\ref{tab:parameters} and their covariances are shown in Fig.~\ref{fig:triangle} \citep[generated using the \textsc{corner} package;][]{Foreman-Mackey16}.  All parameters are well constrained, although there is a degeneracy between inclination angle and parallax.  We measure the angular size of the pulsar orbit on the sky, $a_{\rm NS}/d$, whereas the product $a_{\rm NS}\sin i$ is extremely well constrained from pulsar timing.  Thus, the fitted distance scales inversely with $\sin i$, leading to the degeneracy seen as the slope of the $\pi-i$ contour plot in Fig.~\ref{fig:triangle}.

\subsection{Best-fitting parameters}
Our best-fitting model parameters are given in Table~\ref{tab:parameters}, and for 15 degrees of freedom (22 positional measurements in R.A.\ or Dec., with seven fitted parameters) gave a reduced $\chi^2$ value of 0.54.  While this low reduced $\chi^2$ value could potentially indicate that the systematics have been slightly overestimated, \citet{Andrae10} urge caution in the interpretation of reduced chi-squared values, particularly in the case of non-linear models.  We therefore generated 1000 realisations of the observed data set using the positions given by the best-fitting astrometric parameters on our observation dates, with uncertainties drawn from Gaussian distributions with standard deviations given by the measured positional errors from each epoch.  While the mean reduced $\chi^2$ value was 1.03, the 95\% confidence interval was 0.44--1.88.  Thus, given the astrometric uncertainties from the individual epochs, a reduced $\chi^2$ value as low as 0.54 is not outside the expected range for our non-linear model.  Should these turn out to be overly conservative, our uncertainties on the fitted astrometric and orbital parameters would decrease.  As a test, we tried neglecting the systematic uncertainties, and found very similar fit parameters but with smaller uncertainties, and a higher reduced $\chi^2$ value of 1.2.  Given the above analysis, we adopt the systematic uncertainties described in Section~\ref{sec:systematics} for the remainder of our analysis.

\begin{table*}
	\centering
	\caption{Results of the MCMC fitting to our LBA astrometric data.  Uncertainties represent the 16th and 84th percentiles of the results distribution.  The reference position is given for MJD 57000.  The fitted parallax of $0.38\pm0.05$\,mas corresponds to a distance of $2.6^{+0.4}_{-0.3}$\,kpc (see Fig.~\ref{fig:distance_pdf}). The five orbital elements  derived from long-term pulsar timing \citep[which were held fixed at their best-fitting values from][]{Shannon14} are listed in the second section of the table.}
	\label{tab:parameters}
	\begin{tabular}{lcc}
		\hline
		Parameter & Symbol & Value\\
		\hline
		Reference position in R.A. (J2000) & $\alpha_0$ & 13$^{\rm h}$02$^{\rm m}$47\fs 638337$\pm0.000012$\\
		Reference position in Dec. (J2000) & $\delta_0$ & $-$63$^{\circ}$50$^{\prime}$8.62859$^{\prime\prime}\pm0.00008$\\
		Proper motion in R.A. (mas\,yr$^{-1}$) & $\mu_{\alpha}\cos\delta$ & $-7.01\pm0.03$\\
		Proper motion in Dec. (mas\,yr$^{-1}$) & $\mu_{\delta}$ & $-0.53\pm0.03$ \\
		Parallax (mas) & $\pi$ & $0.38\pm0.05$\\
		Inclination angle (\degr) & $i$ & $154\pm3$\\
		Longitude of the ascending node (\degr\ CCW from N through E) & $\Omega$ & $189\pm2$\\
		\hline
        Orbital period (days) & $P$ & $1236.724526\pm0.000006$\\
        Epoch of periastron (MJD) & $T_0$ & $53071.2447290\pm0.0000007$\\
        Eccentricity & $e$ & $0.86987970\pm0.00000006$\\
        Projected semi-major axis (lt-s) & $a\sin i$ & $1296.27448\pm0.00014$\\
        Argument of periastron & $\omega$ & $138\fdg665013\pm0\fdg000011$\\
        \hline
	\end{tabular}
\end{table*}

\subsubsection{A geometric distance to \psrb}
\label{sec:distance}

\begin{figure}
	\includegraphics[width=\columnwidth]{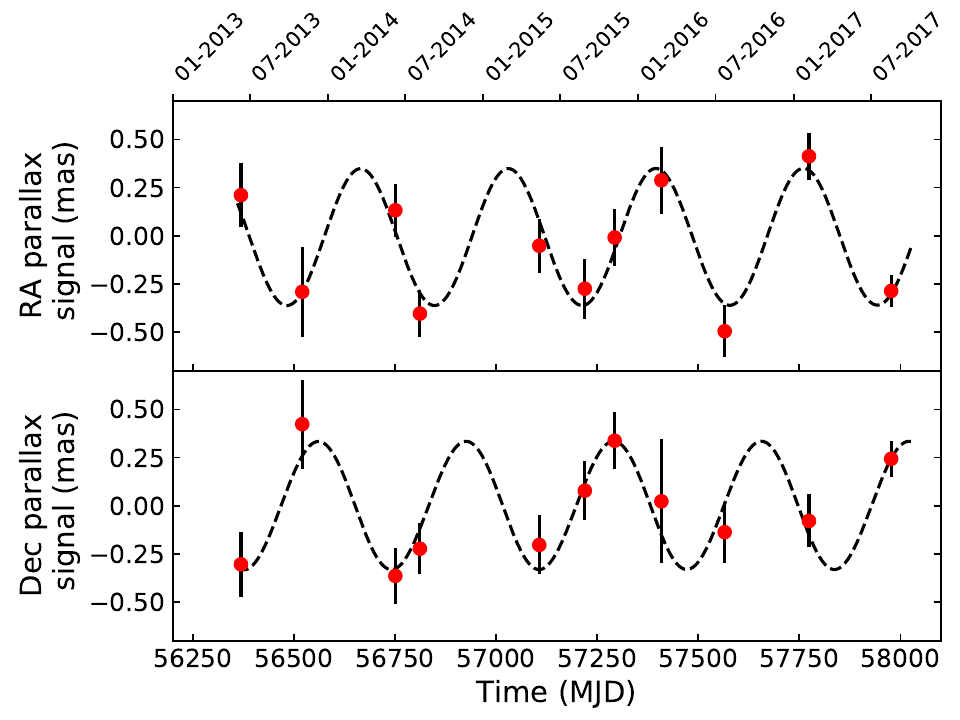}
    \caption{The observed parallax signature of \psrb, in both R.A.\ (top) and Dec.\ (bottom).  The best-fitting proper motion and orbital signatures have been subtracted, leaving only the annual parallax signature.  Inverting the measured parallax gives a model-independent geometric distance of $2.6^{+0.4}_{-0.3}$\,kpc, identical (within uncertainties) to the peak and $1\sigma$ confidence interval calculated using the Bayesian formalism of \citet{Igoshev16}.}
    \label{fig:parallax}
\end{figure}

As shown in Fig.~\ref{fig:parallax}, we achieved good sampling of the parallax extrema in both R.A.\ and Dec., allowing us to constrain the parallax to $0.38\pm0.05$\,mas, placing the source at a direct inversion distance of $2.6^{+0.4}_{-0.3}$\,kpc.  However, we reiterate that the distance is degenerate with the inclination angle of the orbit, as described in Section~\ref{sec:mcmc}.

As originally noted by \citet{Lutz73}, there is a systematic bias in measured trigonometric parallaxes such that for an isotropically-distributed population the measured values tend to be overestimated, owing to the larger volume probed at higher distances. In converting a measured parallax to a distance, it is therefore necessary to account for prior information on the likelihood of measuring that parallax given knowledge of the source population.  \citet*{Verbiest10} analysed this bias specifically for the radio pulsar population, taking into account knowledge of the pulsar spatial distribution in the Galaxy (for this volumetric correction) and the pulsar luminosity function (a luminosity correction).  Their approach was updated by \citet{Verbiest12}, who determined analytical expressions for the probability density functions of both the true parallax and the true distance of a pulsar, given its measured parallax, flux density, Galactic position, and any existing H{\sc i} distance constraints.  However, owing to a slight error in their formulae, we therefore use the corrected formalism of \citet{Igoshev16} to determine the probability density function for the true distance to \psrb.

\begin{figure}
	\centering
	\includegraphics[width=\columnwidth]{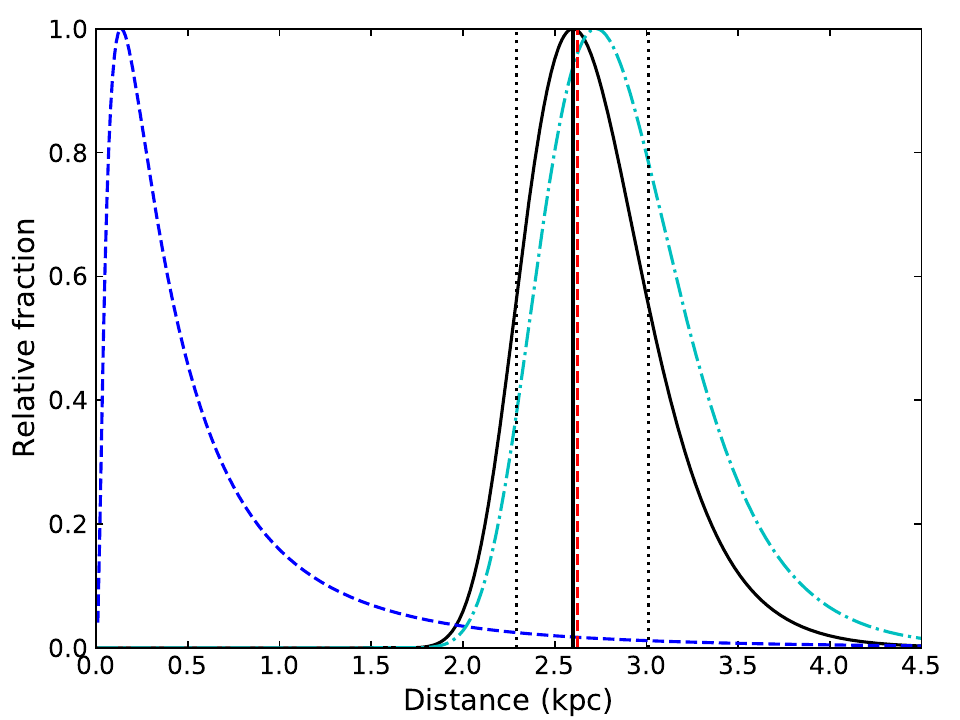}
    \caption{The probability density function (PDF) for the distance to the \psrb/LS\,2883 system, using the formalism of \citet{Igoshev16}. The cyan dot-dashed curve shows the PDF based on the parallax alone, the blue dashed curve shows the lognormal luminosity prior, and the black solid curve shows the overall PDF, with the black vertical line showing the most likely distance of 2.6\,kpc. Black vertical dotted lines show the 68 per cent confidence interval \citep[defined as in][]{Igoshev16} of 2.3--3.0\,kpc, and red vertical dashed line shows the direct inversion distance $1/\pi$, which is almost identical to the most probable distance.}
    \label{fig:distance_pdf}
\end{figure}

We assume as priors a lognormal pulsar luminosity function with mean $\log_{10} L=-1.1$ and standard deviation $\sigma_{\log_{10} L}=0.9$ \citep{Faucher-Giguere06}, and the Galactocentric pulsar distribution of \citet{Verbiest12}.  Using a 1.4-GHz pulsed flux density of 4.2\,mJy \citep[the weighted mean of the measurements tabulated by][]{Johnston05}, and our measured parallax of $0.38\pm0.05$\,mas, we derive the probability density function for the distance to the \psrb/LS\,2883 system shown in Fig.~\ref{fig:distance_pdf}.  This gives a most probable distance of $2.6^{+0.4}_{-0.3}$\,kpc, with the 68 per cent confidence interval calculated following the procedure outlined by \citet{Igoshev16}. We use this value and uncertainty as our best estimate of the distance for the remainder of this paper.

\subsubsection{The binary orbit of \psrb}

\begin{figure}
	\includegraphics[width=\columnwidth]{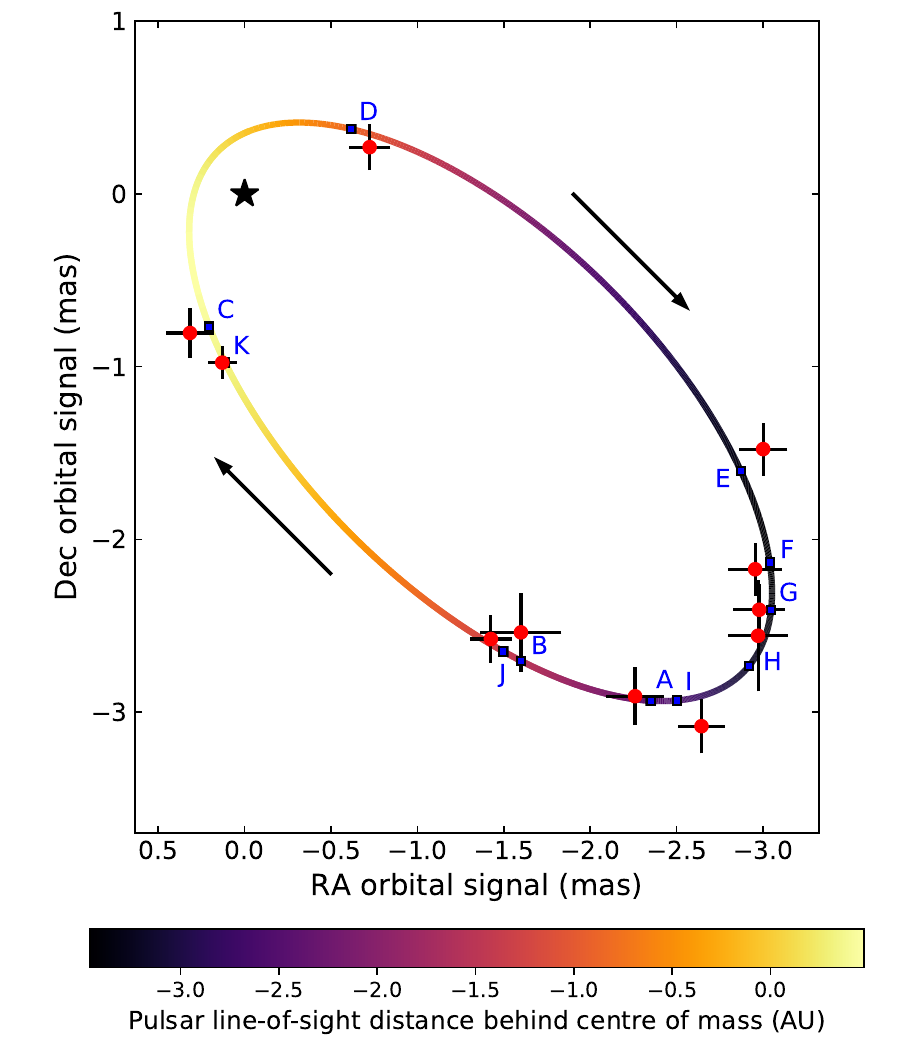}
    \caption{The observed orbital motion of \psrb, after subtracting off the best-fitting proper motion and parallax signatures. The best-fitting orbit is plotted as the solid line, with the colours denoting the pulsar's distance along the line of sight, in  au, relative to the centre of mass of the system.  Yellow denotes when the pulsar is furthest away from the observer, and black when it is closest.  Blue squares show the predicted position at each epoch. Red circles show the measured positions with $1\sigma$ error bars. The centre of mass of the system is shown by the black star.  The pulsar moves clockwise around its orbit, as shown by the arrows.}
    \label{fig:orbit}
\end{figure}

The high eccentricity of the binary orbit \citep[$e=0.86987970(6)$;][]{Shannon14} implies that the pulsar spends the majority of time close to apastron, where the orbital velocity is lowest.  However, three of our epochs (C, D and K) were taken 30--40\,d before or after periastron passage. The pulsar is eclipsed between at least 16\,d before and 15\,d after periastron, which precluded us from taking observations significantly closer in time to periastron passage.  The orbit subtends 4.2\,mas on the sky (Fig.~\ref{fig:orbit}), and its inclination is best constrained by the epochs closest to quadrature (eccentric anomalies of 90\degr\ and 270\degr).  The pulsar moves clockwise around its orbit (as seen on the sky), which is inclined at $154\pm3$\degr\ to the line of sight. The pulsar is closest to Earth around apastron.

\subsection{Transient unpulsed emission}
\label{sec:unpulsed}

Transient, unpulsed emission is known to arise from the system around periastron passage.  At 8.4\,GHz, this has been seen to last from from 15\,d before periastron to 64\,d afterwards \citep{Johnston05}.  Our only epoch falling in this time period is Epoch D (30.0\,d post-periastron).  In that epoch, we detect faint emission during one of the two off-pulse bins, but not in the other.  The absence in the second bin (to a $3\sigma$ upper limit of 0.61\,mJy\,beam$^{-1}$), together with the lack of a significant offset from the fitted pulsar position (angular separation $0.3\pm0.3$\,mas) suggests that the faint off-pulse emission could be due to a change in the pulse profile or dispersion measure close to periastron \citep[e.g.][]{Johnston05} rather than to any unpulsed continuum emission.

The transient unpulsed emission is known to fade significantly at all frequencies between 20 and 30 days after periastron passage, reaching flux densities of 5--10\,mJy at 8.4\,GHz by 28 days post-periastron \citep{Johnston05}.  \citet{Moldon11} and \citet{Chernyakova14} used LBA observations to show that 20--30\,d post-periastron, the extended emission is resolved on a size scale of $\sim50$\,mas at 2.3\,GHz, and has a total flux density of $\sim50$\,mJy.  Owing to the steep spectral index of the unpulsed emission, such emission would be significantly fainter in our 8.4-GHz images, if present.  Furthermore, our higher resolution (a beam size of $2.5\times2.0$\,mas$^2$ in Epoch D) implies that such extended emission would likely be resolved out.  To test for extended unpulsed emission, we therefore re-imaged the off-pulse bin in which no emission was detected in our full-resolution image.  We used only the innermost three antennas (ATCA, Parkes and Mopra; maximum baseline 321\,km).  This gave a synthesised beam size of $42.1\times16.6$\,mas$^2$, ensuring that we were not resolving out any extended emission on the $\sim50$\,mas size scales determined by \citet{Moldon11} and \citet{Chernyakova14}.  We detected an unresolved source of flux density $1.6\pm0.2$\,mJy, showing that the unpulsed emission is indeed present in our data, but that it is resolved out in all our full-resolution images, so we can be confident that it is not affecting our astrometric measurements.

\section{Discussion}
\label{sec:discussion}

The results of our VLBI astrometry, coupled with high-precision pulsar timing \citep{Shannon14}, provide a full orbital solution for the binary system.  Given the 180\degr\ sense ambiguity affecting previous estimates, our derived orbital inclination of $154\pm3$\degr\ to the line of sight is consistent with the $25^{+6}_{-5}$\degr\ derived by \citet{Negueruela11} within the $1\sigma$ uncertainty, but slightly more precise.  Importantly, we constrain for the first time the sense of the orbit, with the pulsar moving clockwise on the plane of the sky.

While our fitted proper motion values differ slightly from the {\it Gaia} DR2 values, the discrepancy (which is most pronounced in Declination) is consistent with expectations from the orbital motion of the stellar component, LS\,2883, which has not been accounted for in the {\it Gaia} DR2 analysis.  Our measured parallax is slightly smaller than the {\it Gaia}-determined value, but consistent within uncertainties.  While the unmodelled orbital motion is likely to affect the {\it Gaia} parallax measurement (as inferred from the poor goodness-of-fit statistic for the {\it Gaia} solution for this system), a full analysis of this effect is beyond the scope of this work.

Our derived distance of $2.6^{+0.4}_{-0.3}$\,kpc is larger than the $2.3\pm0.4$\,kpc estimated by \citet{Negueruela11}, but again consistent within the $1\sigma$ uncertainties.  However, the distance estimate of \citet{Negueruela11} was based on the best previous determination of the distance to Cen OB1, which has recently been revised upwards to $2.6\pm0.5$\,kpc \citep{Corti13}.  Hence our distance remains fully consistent with the \psrb/LS\,2883 system having originated from that OB association \citep[see also][]{Shannon14}.

The increase in the derived distance further increases the inferred gamma-ray luminosity.  To quantify this effect, we use the Galactocentric spatial distribution and lognormal luminosity priors adopted in Section~\ref{sec:distance}, and again adopt the Bayesian formalism of \citet{Igoshev16}.  Our measured parallax coupled with the gamma-ray flux from the 2010 periastron passage reported by \citet{Abdo11} then implies a 68 per cent confidence interval for the gamma-ray luminosity ($>100$\,MeV) of 0.8--1.4$\times10^{36}$\,erg\,s$^{-1}$.  The 100\,MeV--300\,GeV gamma-ray luminosity measured during the 2017 periastron passage was even higher \citep{Johnson17}, increasing the 68 per cent confidence interval for the luminosity to 1.8--3.0$\times10^{36}$\,erg\,s$^{-1}$, formally exceeding the nominal spin-down luminosity of the pulsar of $8.2\times10^{35}$\,erg\,s$^{-1}$. With the caveat that the spin-down luminosity is inherently somewhat uncertain due to the unknown moment of inertia of the pulsar \citep{Taylor93a}, our results would therefore appear to favour gamma-ray production mechanisms involving Doppler boosting \citep[e.g.][]{Dubus10,Tam11,Kong12}.  Finally, our new distance estimate can be used to update the inferred optical luminosity for the stellar companion, LS\,2883 \citep{Negueruela11}, to $L_{\ast} = 2.9^{+1.0}_{-0.6}\times10^{38}$\,erg\,s$^{-1}$.

\subsection{Component masses}

\begin{figure}
	\includegraphics[width=\columnwidth]{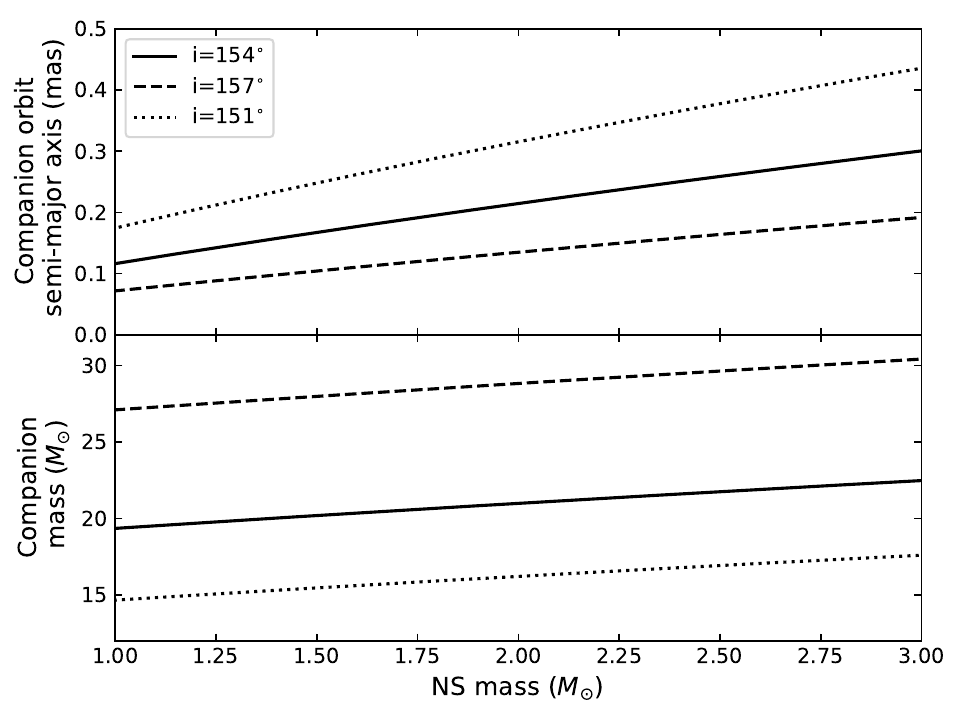}
    \caption{{\it Lower panel}: Constraints on the component masses derived from Kepler's Third Law, plotted for a conservative range of neutron star masses. We use our measured inclination angle, together with the projected semi-major axis of the pulsar orbit and the orbital period derived from pulsar timing \citep{Shannon14}. Solid line shows our best-fitting inclination angle and dotted/dashed lines show the $1\sigma$ upper/lower limits on the inclination.  These inclination constraints imply possible companion masses ranging from 15--31$M_{\odot}$. {\it Upper panel:} Corresponding constraints on the semi-major axis of the companion star, LS\,2883. For our inferred distance of $2.6^{+0.4}_{-0.3}$\,kpc, {\it Gaia} should eventually measure a semi-major axis for the optical companion in the range 70--430\,microarcseconds.}
    \label{fig:masses}
\end{figure}

With a measured inclination angle, we can deproject the semi-major axis of the neutron star orbit derived from pulsar timing, and then use Kepler's Third Law to derive constraints on both the component masses and the semi-major axis of the orbit of the companion star, LS\,2883.  For a conservative range of neutron star masses from 1--3$M_{\odot}$, our $\pm1\sigma$ inclination limits imply companion masses in the range 15--31$M_{\odot}$ (see Figure~\ref{fig:masses}).  Larger inclination angles ($>154$\degr; corresponding to a larger distance) imply a larger neutron star orbit, and hence a larger companion mass for a given neutron star mass.  \citet{Negueruela11} found a companion mass in the range 16--51$M_{\odot}$ when deprojecting the stellar rotational velocity assuming a rotation axis inclined at $35^{\circ}$ to the line of sight.  Our mass constraints, while still relatively broad, significantly reduce the possible parameter space, and push the likely companion mass towards the lower end of the previous range. To further reduce the uncertainty on the inclination angle (and hence on the companion mass), we would need either additional VLBI observations close to quadrature (eccentric anomalies of 90\degr\ or 270\degr) to break the degeneracy between distance and inclination angle, or better distance constraints from {\it Gaia}.

For a given deprojected neutron star orbital semi-major axis, we can also determine $a_{\ast}$, the semi-major axis of the companion star orbit, as a function of neutron star mass, as shown in the upper panel of Figure~\ref{fig:masses}.  Depending on both the neutron star mass and the inclination angle of the binary orbit, we expect a semi-major axis $a_{\ast}$ in the range 70--430\,microarcseconds.  This range is well within the final capabilities of {\it Gaia} given that LS\,2883 has G-magnitude 9.5 \citep{Gaia16a,Carrasco16,Evans17,vanLeeuwen17}. More importantly, the combination of the VLBI-measured orbit of the pulsar with the {\it Gaia}-measured orbit of the companion could eventually enable us to solve for the individual component masses, providing an accurate neutron star mass measurement from the combination of pulsar timing with high-precision optical and radio astrometry.

\subsection{Space velocity and birthplace}

Having improved the precision of the proper motion measurement by an order of magnitude as compared to the previous estimate from {\it Gaia} DR1 \citep{Lindegren16}, we can better constrain the likely birthplace and natal kick of the system.  Using our known distance, then with the systemic radial velocity $\gamma$ we can determine the full three-dimensional space velocity of the system.

\citet{Johnston94} originally suggested that the peak of the H$\alpha$ emission was blueshifted from its rest wavelength by 80\,km\,s$^{-1}$.  However, subsequent works \citep{Negueruela11, vanSoelen16} found negligible systemic velocity relative to the local standard of rest (LSR), and \citet{Shannon14} considered both possibilities in their analysis of the origin of the system.  Closer inspection of the line profiles from \citet{Johnston94} shows the peak of their H$\alpha$ emission line to be at 6562.9\,\AA, very close to the H$\alpha$ rest wavelength in air of 6562.8\,\AA\footnote{See the NIST Atomic Spectra Database, at \url{https://www.nist.gov/pml/atomic-spectra-database}}. The spectra were taken with the UCLES spectrograph on the Anglo-Australian telescope, which sits in air.  The observed line profiles therefore imply a minimal Doppler shift and hence a very small systemic velocity for the \psrb/LS\,2883 system. If instead the observed H$\alpha$ line of \citet{Johnston94} were mistakenly referenced to the rest wavelength in vacuum, at 6564.6\,\AA\ \citep[see][for conversions between air and vacuum wavelengths]{Morton91}, then application of the standard Doppler formula in calculating the velocity shift would give the claimed blueshift of $-82$\,km\,s$^{-1}$.  Since assuming the wrong rest wavelength as above would appear to perfectly explain the existing discrepancy, we therefore assume that the system in fact has close to zero systemic radial velocity.

Having determined all six position-velocity components, we use the transformations of \citet{Johnson87} to deduce the heliocentric Galactic space velocity components, $(U, V, W) = (-72\pm10, -49\pm7, -3\pm1)$\,km\,s$^{-1}$.  Assuming the solar motion, the circular rotation of 240\,km\,s$^{-1}$ and the Galactocentric distance of 8.34\,kpc derived by \citet{Reid14}, this implies a peculiar velocity for the \psrb/LS\,2883 system of $26\pm8$\,km\,s$^{-1}$.  With the characteristic age of the pulsar being $3\times10^5$\,yr \citep{Shannon14}, this implies that it has moved of order 8\,pc from its birthplace, which remains consistent with its having originated in the Cen OB1 association.  \citet{Corti13} determined a distance to Cen OB1 of $2.6\pm0.5$\,kpc, consistent with our distance to \psrb, and measured the proper motion and systemic radial velocity of the association.  Using their measured parameters, we find that Cen OB1 has a peculiar velocity of $23\pm10$\,km\,s$^{-1}$ with respect to its Local Standard of Rest, and that the \psrb/LS\,2883 system has a space velocity of $34\pm13$\,km\,s$^{-1}$ relative to Cen OB1.  This is again consistent with a moderate natal kick for the system \citep[see also][]{Shannon14}.

\subsection{X-ray ejecta}

Using the high spatial resolution of the {\em Chandra} satellite, \citet*{Pavlov11} first detected extended X-ray emission to the south-southwest of the \psrb/LS\,2883 system.  This was later found by \citet{Kargaltsev14} to be moving away from the system at 5 per cent of the speed of light, and interpreted as arising from the pulsar wind ejected from the system close to apastron.  The direction of motion allowed them to suggest an orientation of the orbit on the plane of the sky that is remarkably close to the one that we measure.  We note that in the absence of observational constraints they had to assume the sense of the pulsar's orbital motion.  Our VLBI observations  are uniquely able to measure both the orientation of the orbit on the sky and the sense of rotation, and vindicate the assumptions of \citet{Kargaltsev14}. 

\citet{Pavlov15} added an additional epoch of high-resolution X-ray data to confirm the direction of motion, and found marginal evidence for acceleration.  They interpreted the detected X-ray knot as synchrotron emission from a fragment of the disk around the companion star that moves away in the unshocked, relativistic wind of the pulsar, and hence in the direction away from the companion star.  Alternatively, \citet{Barkov16} suggested that a mixture of pulsar wind and stellar wind was always moving away from the binary in the direction of apastron, and interpreted the emission as non-thermal emission from shocks due to the injection into this flow of stellar wind close to periastron passage.  Regardless, our astrometric observations confirm the consensus that the motion of the X-ray knot is in the direction of apastron.

\subsection{Further modelling}

The well-constrained orbital parameters from our astrometric work should enable a better interpretation of the morphology of the transient unpulsed emission from the \psrb/LS\,2883 system.  \citet*{Moldon12} have shown that the extended emission from the gamma-ray binary LS\,5039 can be modelled as an outflow of relativistic electrons accelerated by the interaction between the stellar wind and the pulsar wind, which trails the pulsar as it moves in its orbit \citep[see][for the relevant model]{Dubus06}.  The predicted morphology depends on the inclination of the orbit and the longitude of the ascending node, so with an accurate knowledge of these parameters, it would be possible to test this model in the \psrb/LS\,2883 system.

Furthermore, with well-constrained astrometric parameters, as well as the improved algorithms that have been developed for modelling pulsar timing data \citep{Lentati14}, it could be possible to improve the estimation of the pulsar timing parameters \citep{Shannon14}.  However, as with the modelling of the extended emission, this is beyond the scope of the current work.

\section{Conclusions}
\label{sec:conclusions}

Using high-precision astrometric observations with the Australian LBA, we have measured the astrometric and orbital parameters of the gamma-ray binary system \psrb/LS\,2883.  From eleven epochs of observation, we measured a parallax of $0.38\pm0.05$\,mas, which we use with accepted priors for the pulsar luminosity and spatial distributions to infer a distance of $2.6^{+0.4}_{-0.3}$\,kpc.  We also measured an inclination angle of the binary orbit of $154\pm3$\degr, which together with existing pulsar timing measurements \citep{Shannon14} implies a companion mass of 15--31$M_{\sun}$.  The pulsar rotates clockwise around its orbit, and the orientation of that orbit on the plane of the sky is consistent with the extended X-ray emission moving away in the direction of apastron \citep{Kargaltsev14,Pavlov15,Barkov16}.  Our measured proper motion implies a space velocity of $34\pm13$\,km\,s$^{-1}$ relative to Cen OB1, and is consistent with the system having been formed in that association and receiving a natal kick on the formation of the neutron star.

\section*{Acknowledgements}

The authors thank Cormac Reynolds and Hayley Bignall for useful discussions regarding LBA calibration, and Andrei Igoshev for his suggestions on the use of measured parallaxes to infer distances and luminosities. The Long Baseline Array is part of the Australia Telescope National Facility which is funded by the Australian Government for operation as a National Facility managed by CSIRO. The Australian SKA Pathfinder is part of the Australia Telescope National Facility which is managed by CSIRO. Operation of ASKAP is funded by the Australian Government with support from the National Collaborative Research Infrastructure Strategy. ASKAP uses the resources of the Pawsey Supercomputing Centre. Establishment of ASKAP, the Murchison Radio-astronomy Observatory and the Pawsey Supercomputing Centre are initiatives of the Australian Government, with support from the Government of Western Australia and the Science and Industry Endowment Fund. We acknowledge the Wajarri Yamatji people as the traditional owners of the Observatory site. This work made use of the Swinburne University of Technology software correlator, developed as part of the Australian Major National Research Facilities Programme. This work was supported by resources provided by the Pawsey Supercomputing Centre with funding from the Australian Government and the Government of Western Australia. DSS34, DSS35, DSS43 and DSS45 are part of the NASA Canberra Deep Space Communication Complex (CDSCC) operated by CSIRO. This work has made use of data from the European Space Agency (ESA)
mission {\it Gaia} (\url{https://www.cosmos.esa.int/gaia}), processed by
the {\it Gaia} Data Processing and Analysis Consortium (DPAC,
\url{https://www.cosmos.esa.int/web/gaia/dpac/consortium}). Funding
for the DPAC has been provided by national institutions, in particular
the institutions participating in the {\it Gaia} Multilateral Agreement.  This research has made use of NASA's Astrophysics Data System, the SIMBAD database, operated at CDS, Strasbourg, France, and Astropy, a community-developed core Python package for Astronomy \citep{astropy13}.  JCAM-J and ATD are the recipients of Australian Research Council Future Fellowships (FT140101082 and FT150100415, respectively). GD acknowledges support from the Centre National d'Etudes Spatiales (CNES). MR and JMP acknowledge support by the Spanish Ministerio de Econom\'ia, Industria y Competitividad (MINEICO/FEDER, UE) under grants FPA2015-69210-C6-2-R, AYA2016-76012-C3-1-P, MDM-2014-0369 of ICCUB (Unidad de Excelencia `Mar\'ia de Maeztu') and the Catalan DEC grant 2017 SGR 643. The National Radio Astronomy Observatory is a facility of the National Science Foundation operated under cooperative agreement by Associated Universities, Inc. SMR is a CIFAR Senior Fellow and receives support from the NSF Physics Frontiers Center award number 1430284.




\bibliographystyle{mnras}
\bibliography{psr_b1259}

\providecommand{\noopsort}[1]{}
\begin{thebibliography}{}
\makeatletter
\relax
\def\mn@urlcharsother{\let\do\@makeother \do\$\do\&\do\#\do\^\do\_\do\%\do\~}
\def\mn@doi{\begingroup\mn@urlcharsother \@ifnextchar [ {\mn@doi@}
  {\mn@doi@[]}}
\def\mn@doi@[#1]#2{\def\@tempa{#1}\ifx\@tempa\@empty \href
  {http://dx.doi.org/#2} {doi:#2}\else \href {http://dx.doi.org/#2} {#1}\fi
  \endgroup}
\def\mn@eprint#1#2{\mn@eprint@#1:#2::\@nil}
\def\mn@eprint@arXiv#1{\href {http://arxiv.org/abs/#1} {{\tt arXiv:#1}}}
\def\mn@eprint@dblp#1{\href {http://dblp.uni-trier.de/rec/bibtex/#1.xml}
  {dblp:#1}}
\def\mn@eprint@#1:#2:#3:#4\@nil{\def\@tempa {#1}\def\@tempb {#2}\def\@tempc
  {#3}\ifx \@tempc \@empty \let \@tempc \@tempb \let \@tempb \@tempa \fi \ifx
  \@tempb \@empty \def\@tempb {arXiv}\fi \@ifundefined
  {mn@eprint@\@tempb}{\@tempb:\@tempc}{\expandafter \expandafter \csname
  mn@eprint@\@tempb\endcsname \expandafter{\@tempc}}}

\bibitem[\protect\citeauthoryear{{Abdo} et~al.,}{{Abdo} et~al.}{2011}]{Abdo11}
{Abdo} A.~A.,  et~al., 2011, \mn@doi [\apjl] {10.1088/2041-8205/736/1/L11},
  \href {http://adsabs.harvard.edu/abs/2011ApJ...736L..11A} {736, L11}

\bibitem[\protect\citeauthoryear{{Aharonian} et~al.,}{{Aharonian}
  et~al.}{2005}]{Aharonian05}
{Aharonian} F.,  et~al., 2005, \mn@doi [\aap] {10.1051/0004-6361:20052983},
  \href {http://adsabs.harvard.edu/abs/2005A%26A...442....1A} {442, 1}

\bibitem[\protect\citeauthoryear{{Aharonian} et~al.,}{{Aharonian}
  et~al.}{2009}]{Aharonian09}
{Aharonian} F.,  et~al., 2009, \mn@doi [\aap] {10.1051/0004-6361/200912339},
  \href {http://adsabs.harvard.edu/abs/2009A%26A...507..389A} {507, 389}

\bibitem[\protect\citeauthoryear{{Andrae}, {Schulze-Hartung}  \&
  {Melchior}}{{Andrae} et~al.}{2010}]{Andrae10}
{Andrae} R.,  {Schulze-Hartung} T.,   {Melchior} P.,  2010, preprint, \href
  {http://adsabs.harvard.edu/abs/2010arXiv1012.3754A} {} (\mn@eprint {arXiv}
  {1012.3754})

\bibitem[\protect\citeauthoryear{{Asaki} et~al.,}{{Asaki}
  et~al.}{2007}]{Asaki07}
{Asaki} Y.,  et~al., 2007, \mn@doi [\pasj] {10.1093/pasj/59.2.397}, \href
  {http://adsabs.harvard.edu/abs/2007PASJ...59..397A} {59, 397}

\bibitem[\protect\citeauthoryear{{Astropy Collaboration} et~al.,}{{Astropy
  Collaboration} et~al.}{2013}]{astropy13}
{Astropy Collaboration} et~al., 2013, \mn@doi [\aap]
  {10.1051/0004-6361/201322068}, \href
  {http://adsabs.harvard.edu/abs/2013A%26A...558A..33A} {558, A33}

\bibitem[\protect\citeauthoryear{{Barkov} \& {Bosch-Ramon}}{{Barkov} \&
  {Bosch-Ramon}}{2016}]{Barkov16}
{Barkov} M.~V.,  {Bosch-Ramon} V.,  2016, \mn@doi [\mnras]
  {10.1093/mnrasl/slv171}, \href
  {http://adsabs.harvard.edu/abs/2016MNRAS.456L..64B} {456, L64}

\bibitem[\protect\citeauthoryear{{Barron}, {Kaplan}, {Bangert}, {Bartlett},
  {Puatua}, {Harris}  \& {Barrett}}{{Barron} et~al.}{2011}]{Barron11}
{Barron} E.~G.,  {Kaplan} G.~H.,  {Bangert} J.,  {Bartlett} J.~L.,  {Puatua}
  W.,  {Harris} W.,   {Barrett} P.,  2011, in American Astronomical Society
  Meeting Abstracts \#217. p. 344.14

\bibitem[\protect\citeauthoryear{{Carrasco} et~al.,}{{Carrasco}
  et~al.}{2016}]{Carrasco16}
{Carrasco} J.~M.,  et~al., 2016, \mn@doi [\aap] {10.1051/0004-6361/201629235},
  \href {http://adsabs.harvard.edu/abs/2016A%26A...595A...7C} {595, A7}

\bibitem[\protect\citeauthoryear{{Chernyakova} et~al.,}{{Chernyakova}
  et~al.}{2014}]{Chernyakova14}
{Chernyakova} M.,  et~al., 2014, \mn@doi [\mnras] {10.1093/mnras/stu021}, \href
  {http://adsabs.harvard.edu/abs/2014MNRAS.439..432C} {439, 432}

\bibitem[\protect\citeauthoryear{{Chernyakova} et~al.,}{{Chernyakova}
  et~al.}{2015}]{Chernyakova15}
{Chernyakova} M.,  et~al., 2015, \mn@doi [\mnras] {10.1093/mnras/stv1988},
  \href {http://adsabs.harvard.edu/abs/2015MNRAS.454.1358C} {454, 1358}

\bibitem[\protect\citeauthoryear{{Corti} \& {Orellana}}{{Corti} \&
  {Orellana}}{2013}]{Corti13}
{Corti} M.~A.,  {Orellana} R.~B.,  2013, \mn@doi [\aap]
  {10.1051/0004-6361/201220743}, \href
  {http://adsabs.harvard.edu/abs/2013A%26A...553A.108C} {553, A108}

\bibitem[\protect\citeauthoryear{{Deller}, {Tingay}, {Bailes}  \&
  {West}}{{Deller} et~al.}{2007}]{Deller07}
{Deller} A.~T.,  {Tingay} S.~J.,  {Bailes} M.,   {West} C.,  2007, \mn@doi
  [\pasp] {10.1086/513572}, \href
  {http://adsabs.harvard.edu/abs/2007PASP..119..318D} {119, 318}

\bibitem[\protect\citeauthoryear{{Deller}, {Tingay}, {Bailes}  \&
  {Reynolds}}{{Deller} et~al.}{2009}]{Deller09}
{Deller} A.~T.,  {Tingay} S.~J.,  {Bailes} M.,   {Reynolds} J.~E.,  2009,
  \mn@doi [\apj] {10.1088/0004-637X/701/2/1243}, \href
  {http://adsabs.harvard.edu/abs/2009ApJ...701.1243D} {701, 1243}

\bibitem[\protect\citeauthoryear{{Deller}, {Boyles}, {Lorimer}, {Kaspi},
  {McLaughlin}, {Ransom}, {Stairs}  \& {Stovall}}{{Deller}
  et~al.}{2013}]{Deller13}
{Deller} A.~T.,  {Boyles} J.,  {Lorimer} D.~R.,  {Kaspi} V.~M.,  {McLaughlin}
  M.~A.,  {Ransom} S.,  {Stairs} I.~H.,   {Stovall} K.,  2013, \mn@doi [\apj]
  {10.1088/0004-637X/770/2/145}, \href
  {http://adsabs.harvard.edu/abs/2013ApJ...770..145D} {770, 145}

\bibitem[\protect\citeauthoryear{{Deller} et~al.,}{{Deller}
  et~al.}{2016}]{Deller16}
{Deller} A.~T.,  et~al., 2016, \mn@doi [\apj] {10.3847/0004-637X/828/1/8},
  \href {http://adsabs.harvard.edu/abs/2016ApJ...828....8D} {828, 8}

\bibitem[\protect\citeauthoryear{{Dubus}}{{Dubus}}{2006}]{Dubus06}
{Dubus} G.,  2006, \mn@doi [\aap] {10.1051/0004-6361:20054779}, \href
  {http://adsabs.harvard.edu/abs/2006A%26A...456..801D} {456, 801}

\bibitem[\protect\citeauthoryear{{Dubus}}{{Dubus}}{2013}]{Dubus13}
{Dubus} G.,  2013, \mn@doi [\aapr] {10.1007/s00159-013-0064-5}, \href
  {http://adsabs.harvard.edu/abs/2013A%26ARv..21...64D} {21, 64}

\bibitem[\protect\citeauthoryear{{Dubus} \& {Cerutti}}{{Dubus} \&
  {Cerutti}}{2013}]{Dubus13a}
{Dubus} G.,  {Cerutti} B.,  2013, \mn@doi [\aap] {10.1051/0004-6361/201321741},
  \href {http://adsabs.harvard.edu/abs/2013A%26A...557A.127D} {557, A127}

\bibitem[\protect\citeauthoryear{{Dubus}, {Cerutti}  \& {Henri}}{{Dubus}
  et~al.}{2010}]{Dubus10}
{Dubus} G.,  {Cerutti} B.,   {Henri} G.,  2010, \mn@doi [\aap]
  {10.1051/0004-6361/201014023}, \href
  {http://adsabs.harvard.edu/abs/2010A%26A...516A..18D} {516, A18}

\bibitem[\protect\citeauthoryear{{Dubus}, {Guillard}, {Petrucci}  \&
  {Martin}}{{Dubus} et~al.}{2017}]{Dubus17}
{Dubus} G.,  {Guillard} N.,  {Petrucci} P.-O.,   {Martin} P.,  2017, \mn@doi
  [\aap] {10.1051/0004-6361/201731084}, \href
  {http://adsabs.harvard.edu/abs/2017A%26A...608A..59D} {608, A59}

\bibitem[\protect\citeauthoryear{{Evans} et~al.,}{{Evans}
  et~al.}{2017}]{Evans17}
{Evans} D.~W.,  et~al., 2017, \mn@doi [\aap] {10.1051/0004-6361/201629241},
  \href {http://adsabs.harvard.edu/abs/2017A%26A...600A..51E} {600, A51}

\bibitem[\protect\citeauthoryear{{Faucher-Gigu{\`e}re} \&
  {Kaspi}}{{Faucher-Gigu{\`e}re} \& {Kaspi}}{2006}]{Faucher-Giguere06}
{Faucher-Gigu{\`e}re} C.-A.,  {Kaspi} V.~M.,  2006, \mn@doi [\apj]
  {10.1086/501516}, \href {http://adsabs.harvard.edu/abs/2006ApJ...643..332F}
  {643, 332}

\bibitem[\protect\citeauthoryear{{Foreman-Mackey}}{{Foreman-Mackey}}{2016}]{Foreman-Mackey16}
{Foreman-Mackey} D.,  2016, \mn@doi [The Journal of Open Source Software]
  {10.21105/joss.00024}, 24

\bibitem[\protect\citeauthoryear{{Foreman-Mackey}, {Hogg}, {Lang}  \&
  {Goodman}}{{Foreman-Mackey} et~al.}{2013}]{Foreman-Mackey13}
{Foreman-Mackey} D.,  {Hogg} D.~W.,  {Lang} D.,   {Goodman} J.,  2013, \mn@doi
  [\pasp] {10.1086/670067}, \href
  {http://adsabs.harvard.edu/abs/2013PASP..125..306F} {125, 306}

\bibitem[\protect\citeauthoryear{{Gaia Collaboration} et~al.,}{{Gaia
  Collaboration} et~al.}{2016a}]{Gaia16}
{Gaia Collaboration} et~al., 2016a, \mn@doi [\aap]
  {10.1051/0004-6361/201629272}, \href
  {http://adsabs.harvard.edu/abs/2016A%26A...595A...1G} {595, A1}

\bibitem[\protect\citeauthoryear{{Gaia Collaboration} et~al.,}{{Gaia
  Collaboration} et~al.}{2016b}]{Gaia16a}
{Gaia Collaboration} et~al., 2016b, \mn@doi [\aap]
  {10.1051/0004-6361/201629512}, \href
  {http://adsabs.harvard.edu/abs/2016A%26A...595A...2G} {595, A2}

\bibitem[\protect\citeauthoryear{{Georgelin} \& {Georgelin}}{{Georgelin} \&
  {Georgelin}}{1976}]{Georgelin76}
{Georgelin} Y.~M.,  {Georgelin} Y.~P.,  1976, \aap, \href
  {http://adsabs.harvard.edu/abs/1976A%26A....49...57G} {49, 57}

\bibitem[\protect\citeauthoryear{{Greisen}}{{Greisen}}{2003}]{Greisen03}
{Greisen} E.~W.,  2003, in {Heck} A.,  ed.,  Astrophysics and Space Science
  Library Vol. 285, Information Handling in Astronomy - Historical Vistas.
  p.~109, \mn@doi{10.1007/0-306-48080-8_7}

\bibitem[\protect\citeauthoryear{{Igoshev}, {Verbunt}  \& {Cator}}{{Igoshev}
  et~al.}{2016}]{Igoshev16}
{Igoshev} A.,  {Verbunt} F.,   {Cator} E.,  2016, \mn@doi [\aap]
  {10.1051/0004-6361/201527471}, \href
  {http://adsabs.harvard.edu/abs/2016A%26A...591A.123I} {591, A123}

\bibitem[\protect\citeauthoryear{{Johnson} \& {Soderblom}}{{Johnson} \&
  {Soderblom}}{1987}]{Johnson87}
{Johnson} D.~R.~H.,  {Soderblom} D.~R.,  1987, \mn@doi [\aj] {10.1086/114370},
  \href {http://adsabs.harvard.edu/abs/1987AJ.....93..864J} {93, 864}

\bibitem[\protect\citeauthoryear{{Johnson}, {Wood}, {Ray}, {Kerr}  \&
  {Cheung}}{{Johnson} et~al.}{2017}]{Johnson17}
{Johnson} T.~J.,  {Wood} K.~S.,  {Ray} P.~S.,  {Kerr} M.~T.,   {Cheung} C.~C.,
  2017, The Astronomer's Telegram, \href
  {http://adsabs.harvard.edu/abs/2017ATel10972....1J} {10972}

\bibitem[\protect\citeauthoryear{{Johnston}, {Manchester}, {Lyne}, {Bailes},
  {Kaspi}, {Qiao}  \& {D'Amico}}{{Johnston} et~al.}{1992}]{Johnston92}
{Johnston} S.,  {Manchester} R.~N.,  {Lyne} A.~G.,  {Bailes} M.,  {Kaspi}
  V.~M.,  {Qiao} G.,   {D'Amico} N.,  1992, \mn@doi [\apjl] {10.1086/186300},
  \href {http://adsabs.harvard.edu/abs/1992ApJ...387L..37J} {387, L37}

\bibitem[\protect\citeauthoryear{{Johnston}, {Manchester}, {Lyne}, {Nicastro}
  \& {Spyromilio}}{{Johnston} et~al.}{1994}]{Johnston94}
{Johnston} S.,  {Manchester} R.~N.,  {Lyne} A.~G.,  {Nicastro} L.,
  {Spyromilio} J.,  1994, \mn@doi [\mnras] {10.1093/mnras/268.2.430}, \href
  {http://adsabs.harvard.edu/abs/1994MNRAS.268..430J} {268, 430}

\bibitem[\protect\citeauthoryear{{Johnston}, {Ball}, {Wang}  \&
  {Manchester}}{{Johnston} et~al.}{2005}]{Johnston05}
{Johnston} S.,  {Ball} L.,  {Wang} N.,   {Manchester} R.~N.,  2005, \mn@doi
  [\mnras] {10.1111/j.1365-2966.2005.08854.x}, \href
  {http://adsabs.harvard.edu/abs/2005MNRAS.358.1069J} {358, 1069}

\bibitem[\protect\citeauthoryear{{Kargaltsev}, {Pavlov}, {Durant}, {Volkov}  \&
  {Hare}}{{Kargaltsev} et~al.}{2014}]{Kargaltsev14}
{Kargaltsev} O.,  {Pavlov} G.~G.,  {Durant} M.,  {Volkov} I.,   {Hare} J.,
  2014, \mn@doi [\apj] {10.1088/0004-637X/784/2/124}, \href
  {http://adsabs.harvard.edu/abs/2014ApJ...784..124K} {784, 124}

\bibitem[\protect\citeauthoryear{{Khangulyan}, {Aharonian}, {Bogovalov}  \&
  {Rib{\'o}}}{{Khangulyan} et~al.}{2011}]{Khangulyan11}
{Khangulyan} D.,  {Aharonian} F.~A.,  {Bogovalov} S.~V.,   {Rib{\'o}} M.,
  2011, \mn@doi [\apj] {10.1088/0004-637X/742/2/98}, \href
  {http://adsabs.harvard.edu/abs/2011ApJ...742...98K} {742, 98}

\bibitem[\protect\citeauthoryear{{Khangulyan}, {Aharonian}, {Bogovalov}  \&
  {Rib{\'o}}}{{Khangulyan} et~al.}{2012}]{Khangulyan12}
{Khangulyan} D.,  {Aharonian} F.~A.,  {Bogovalov} S.~V.,   {Rib{\'o}} M.,
  2012, \mn@doi [\apjl] {10.1088/2041-8205/752/1/L17}, \href
  {http://adsabs.harvard.edu/abs/2012ApJ...752L..17K} {752, L17}

\bibitem[\protect\citeauthoryear{{Kirk}, {Ball}  \& {Skj{\ae}raasen}}{{Kirk}
  et~al.}{1999}]{Kirk99}
{Kirk} J.~G.,  {Ball} L.,   {Skj{\ae}raasen} O.,  1999, \mn@doi [Astroparticle
  Physics] {10.1016/S0927-6505(98)00041-3}, \href
  {http://adsabs.harvard.edu/abs/1999APh....10...31K} {10, 31}

\bibitem[\protect\citeauthoryear{{Kong}, {Cheng}  \& {Huang}}{{Kong}
  et~al.}{2012}]{Kong12}
{Kong} S.~W.,  {Cheng} K.~S.,   {Huang} Y.~F.,  2012, \mn@doi [\apj]
  {10.1088/0004-637X/753/2/127}, \href
  {http://adsabs.harvard.edu/abs/2012ApJ...753..127K} {753, 127}

\bibitem[\protect\citeauthoryear{{\noopsort{Leeuwen}}{van Leeuwen}
  et~al.,}{{\noopsort{Leeuwen}}{van Leeuwen} et~al.}{2017}]{vanLeeuwen17}
{\noopsort{Leeuwen}}{van Leeuwen} F.,  et~al., 2017, \mn@doi [\aap]
  {10.1051/0004-6361/201630064}, \href
  {http://adsabs.harvard.edu/abs/2017A%26A...599A..32V} {599, A32}

\bibitem[\protect\citeauthoryear{{Lentati}, {Alexander}, {Hobson}, {Feroz},
  {van Haasteren}, {Lee}  \& {Shannon}}{{Lentati} et~al.}{2014}]{Lentati14}
{Lentati} L.,  {Alexander} P.,  {Hobson} M.~P.,  {Feroz} F.,  {van Haasteren}
  R.,  {Lee} K.~J.,   {Shannon} R.~M.,  2014, \mn@doi [\mnras]
  {10.1093/mnras/stt2122}, \href
  {http://adsabs.harvard.edu/abs/2014MNRAS.437.3004L} {437, 3004}

\bibitem[\protect\citeauthoryear{{Lindegren} et~al.,}{{Lindegren}
  et~al.}{2016}]{Lindegren16}
{Lindegren} L.,  et~al., 2016, \mn@doi [\aap] {10.1051/0004-6361/201628714},
  \href {http://adsabs.harvard.edu/abs/2016A%26A...595A...4L} {595, A4}

\bibitem[\protect\citeauthoryear{{Lindegren} et~al.,}{{Lindegren}
  et~al.}{2018}]{Lindegren18}
{Lindegren} L.,  et~al., 2018, preprint, \href
  {http://adsabs.harvard.edu/abs/2018arXiv180409366L} {} (\mn@eprint {arXiv}
  {1804.09366})

\bibitem[\protect\citeauthoryear{{Loinard}, {Mioduszewski}, {Rodr{\'{\i}}guez},
  {Gonz{\'a}lez}, {Rodr{\'{\i}}guez}  \& {Torres}}{{Loinard}
  et~al.}{2005}]{Loinard05}
{Loinard} L.,  {Mioduszewski} A.~J.,  {Rodr{\'{\i}}guez} L.~F.,  {Gonz{\'a}lez}
  R.~A.,  {Rodr{\'{\i}}guez} M.~I.,   {Torres} R.~M.,  2005, \mn@doi [\apjl]
  {10.1086/428349}, \href {http://adsabs.harvard.edu/abs/2005ApJ...619L.179L}
  {619, L179}

\bibitem[\protect\citeauthoryear{{Luri} et~al.,}{{Luri} et~al.}{2018}]{Luri18}
{Luri} X.,  et~al., 2018, preprint, \href
  {http://adsabs.harvard.edu/abs/2018arXiv180409376L} {} (\mn@eprint {arXiv}
  {1804.09376})

\bibitem[\protect\citeauthoryear{{Lutz} \& {Kelker}}{{Lutz} \&
  {Kelker}}{1973}]{Lutz73}
{Lutz} T.~E.,  {Kelker} D.~H.,  1973, \mn@doi [\pasp] {10.1086/129506}, \href
  {http://adsabs.harvard.edu/abs/1973PASP...85..573L} {85, 573}

\bibitem[\protect\citeauthoryear{{Lyne} \& {Lorimer}}{{Lyne} \&
  {Lorimer}}{1994}]{Lyne94}
{Lyne} A.~G.,  {Lorimer} D.~R.,  1994, \mn@doi [\nat] {10.1038/369127a0}, \href
  {http://adsabs.harvard.edu/abs/1994Natur.369..127L} {369, 127}

\bibitem[\protect\citeauthoryear{{Lyne}, {Manchester}  \& {Taylor}}{{Lyne}
  et~al.}{1985}]{Lyne85}
{Lyne} A.~G.,  {Manchester} R.~N.,   {Taylor} J.~H.,  1985, \mn@doi [\mnras]
  {10.1093/mnras/213.3.613}, \href
  {http://adsabs.harvard.edu/abs/1985MNRAS.213..613L} {213, 613}

\bibitem[\protect\citeauthoryear{{Lyne}, {Stappers}, {Keith}, {Ray}, {Kerr},
  {Camilo}  \& {Johnson}}{{Lyne} et~al.}{2015}]{Lyne15}
{Lyne} A.~G.,  {Stappers} B.~W.,  {Keith} M.~J.,  {Ray} P.~S.,  {Kerr} M.,
  {Camilo} F.,   {Johnson} T.~J.,  2015, \mn@doi [\mnras]
  {10.1093/mnras/stv236}, \href
  {http://adsabs.harvard.edu/abs/2015MNRAS.451..581L} {451, 581}

\bibitem[\protect\citeauthoryear{{Manchester} \& {Johnston}}{{Manchester} \&
  {Johnston}}{1995}]{Manchester95a}
{Manchester} R.~N.,  {Johnston} S.,  1995, \mn@doi [\apjl] {10.1086/187791},
  \href {http://adsabs.harvard.edu/abs/1995ApJ...441L..65M} {441, L65}

\bibitem[\protect\citeauthoryear{{Manchester}, {Johnston}, {Lyne}, {D'Amico},
  {Bailes}  \& {Nicastro}}{{Manchester} et~al.}{1995}]{Manchester95}
{Manchester} R.~N.,  {Johnston} S.,  {Lyne} A.~G.,  {D'Amico} N.,  {Bailes} M.,
    {Nicastro} L.,  1995, \mn@doi [\apjl] {10.1086/187908}, \href
  {http://adsabs.harvard.edu/abs/1995ApJ...445L.137M} {445, L137}

\bibitem[\protect\citeauthoryear{{Melatos}, {Johnston}  \& {Melrose}}{{Melatos}
  et~al.}{1995}]{Melatos95}
{Melatos} A.,  {Johnston} S.,   {Melrose} D.~B.,  1995, \mn@doi [\mnras]
  {10.1093/mnras/275.2.381}, \href
  {http://adsabs.harvard.edu/abs/1995MNRAS.275..381M} {275, 381}

\bibitem[\protect\citeauthoryear{{Miller-Jones}, {Jonker}, {Dhawan}, {Brisken},
  {Rupen}, {Nelemans}  \& {Gallo}}{{Miller-Jones}
  et~al.}{2009}]{Miller-Jones09}
{Miller-Jones} J.~C.~A.,  {Jonker} P.~G.,  {Dhawan} V.,  {Brisken} W.,  {Rupen}
  M.~P.,  {Nelemans} G.,   {Gallo} E.,  2009, \mn@doi [\apjl]
  {10.1088/0004-637X/706/2/L230}, \href
  {http://adsabs.harvard.edu/abs/2009ApJ...706L.230M} {706, L230}

\bibitem[\protect\citeauthoryear{{Mold{\'o}n}, {Johnston}, {Rib{\'o}},
  {Paredes}  \& {Deller}}{{Mold{\'o}n} et~al.}{2011}]{Moldon11}
{Mold{\'o}n} J.,  {Johnston} S.,  {Rib{\'o}} M.,  {Paredes} J.~M.,   {Deller}
  A.~T.,  2011, \mn@doi [\apjl] {10.1088/2041-8205/732/1/L10}, \href
  {http://adsabs.harvard.edu/abs/2011ApJ...732L..10M} {732, L10}

\bibitem[\protect\citeauthoryear{{Mold{\'o}n}, {Rib{\'o}}  \&
  {Paredes}}{{Mold{\'o}n} et~al.}{2012}]{Moldon12}
{Mold{\'o}n} J.,  {Rib{\'o}} M.,   {Paredes} J.~M.,  2012, \mn@doi [\aap]
  {10.1051/0004-6361/201219553}, \href
  {http://adsabs.harvard.edu/abs/2012A%26A...548A.103M} {548, A103}

\bibitem[\protect\citeauthoryear{{Morton}}{{Morton}}{1991}]{Morton91}
{Morton} D.~C.,  1991, \mn@doi [\apjs] {10.1086/191601}, \href
  {http://adsabs.harvard.edu/abs/1991ApJS...77..119M} {77, 119}

\bibitem[\protect\citeauthoryear{{Negueruela}, {Rib{\'o}}, {Herrero},
  {Lorenzo}, {Khangulyan}  \& {Aharonian}}{{Negueruela}
  et~al.}{2011}]{Negueruela11}
{Negueruela} I.,  {Rib{\'o}} M.,  {Herrero} A.,  {Lorenzo} J.,  {Khangulyan}
  D.,   {Aharonian} F.~A.,  2011, \mn@doi [\apjl]
  {10.1088/2041-8205/732/1/L11}, \href
  {http://adsabs.harvard.edu/abs/2011ApJ...732L..11N} {732, L11}

\bibitem[\protect\citeauthoryear{{Pavlov}, {Chang}  \& {Kargaltsev}}{{Pavlov}
  et~al.}{2011}]{Pavlov11}
{Pavlov} G.~G.,  {Chang} C.,   {Kargaltsev} O.,  2011, \mn@doi [\apj]
  {10.1088/0004-637X/730/1/2}, \href
  {http://adsabs.harvard.edu/abs/2011ApJ...730....2P} {730, 2}

\bibitem[\protect\citeauthoryear{{Pavlov}, {Hare}, {Kargaltsev}, {Rangelov}  \&
  {Durant}}{{Pavlov} et~al.}{2015}]{Pavlov15}
{Pavlov} G.~G.,  {Hare} J.,  {Kargaltsev} O.,  {Rangelov} B.,   {Durant} M.,
  2015, \mn@doi [\apj] {10.1088/0004-637X/806/2/192}, \href
  {http://adsabs.harvard.edu/abs/2015ApJ...806..192P} {806, 192}

\bibitem[\protect\citeauthoryear{{P{\'e}tri} \& {Dubus}}{{P{\'e}tri} \&
  {Dubus}}{2011}]{Petri11}
{P{\'e}tri} J.,  {Dubus} G.,  2011, \mn@doi [\mnras]
  {10.1111/j.1365-2966.2011.19295.x}, \href
  {http://adsabs.harvard.edu/abs/2011MNRAS.417..532P} {417, 532}

\bibitem[\protect\citeauthoryear{{Petrov}, {Phillips}, {Bertarini}, {Murphy}
  \& {Sadler}}{{Petrov} et~al.}{2011}]{Petrov11}
{Petrov} L.,  {Phillips} C.,  {Bertarini} A.,  {Murphy} T.,   {Sadler} E.~M.,
  2011, \mn@doi [\mnras] {10.1111/j.1365-2966.2011.18570.x}, \href
  {https://ui.adsabs.harvard.edu/#abs/2011MNRAS.414.2528P} {414, 2528}

\bibitem[\protect\citeauthoryear{{Ransom}}{{Ransom}}{2001}]{Ransom01}
{Ransom} S.~M.,  2001, PhD thesis, Harvard University

\bibitem[\protect\citeauthoryear{{Reid} et~al.,}{{Reid} et~al.}{2014}]{Reid14}
{Reid} M.~J.,  et~al., 2014, \mn@doi [\apj] {10.1088/0004-637X/783/2/130},
  \href {http://adsabs.harvard.edu/abs/2014ApJ...783..130R} {783, 130}

\bibitem[\protect\citeauthoryear{{Rivinius}, {Carciofi}  \&
  {Martayan}}{{Rivinius} et~al.}{2013}]{Rivinius13}
{Rivinius} T.,  {Carciofi} A.~C.,   {Martayan} C.,  2013, \mn@doi [\aapr]
  {10.1007/s00159-013-0069-0}, \href
  {http://adsabs.harvard.edu/abs/2013A%26ARv..21...69R} {21, 69}

\bibitem[\protect\citeauthoryear{{Shannon}, {Johnston}  \&
  {Manchester}}{{Shannon} et~al.}{2014}]{Shannon14}
{Shannon} R.~M.,  {Johnston} S.,   {Manchester} R.~N.,  2014, \mn@doi [\mnras]
  {10.1093/mnras/stt2123}, \href
  {http://adsabs.harvard.edu/abs/2014MNRAS.437.3255S} {437, 3255}

\bibitem[\protect\citeauthoryear{{\noopsort{Soelen}}{van Soelen},
  {V{\"a}is{\"a}nen}, {Odendaal}, {Klindt}, {Sushch}  \&
  {Meintjes}}{{\noopsort{Soelen}}{van Soelen} et~al.}{2016}]{vanSoelen16}
{\noopsort{Soelen}}{van Soelen} B.,  {V{\"a}is{\"a}nen} P.,  {Odendaal} A.,
  {Klindt} L.,  {Sushch} I.,   {Meintjes} P.~J.,  2016, \mn@doi [\mnras]
  {10.1093/mnras/stv2576}, \href
  {http://adsabs.harvard.edu/abs/2016MNRAS.455.3674V} {455, 3674}

\bibitem[\protect\citeauthoryear{{Tam}, {Huang}, {Takata}, {Hui}, {Kong}  \&
  {Cheng}}{{Tam} et~al.}{2011}]{Tam11}
{Tam} P.~H.~T.,  {Huang} R.~H.~H.,  {Takata} J.,  {Hui} C.~Y.,  {Kong}
  A.~K.~H.,   {Cheng} K.~S.,  2011, \mn@doi [\apjl]
  {10.1088/2041-8205/736/1/L10}, \href
  {http://adsabs.harvard.edu/abs/2011ApJ...736L..10T} {736, L10}

\bibitem[\protect\citeauthoryear{{Tavani}, {Arons}  \& {Kaspi}}{{Tavani}
  et~al.}{1994}]{Tavani94}
{Tavani} M.,  {Arons} J.,   {Kaspi} V.~M.,  1994, \mn@doi [\apjl]
  {10.1086/187542}, \href {http://adsabs.harvard.edu/abs/1994ApJ...433L..37T}
  {433, L37}

\bibitem[\protect\citeauthoryear{{Taylor} \& {Cordes}}{{Taylor} \&
  {Cordes}}{1993}]{Taylor93}
{Taylor} J.~H.,  {Cordes} J.~M.,  1993, \mn@doi [\apj] {10.1086/172870}, \href
  {http://adsabs.harvard.edu/abs/1993ApJ...411..674T} {411, 674}

\bibitem[\protect\citeauthoryear{{Taylor}, {Manchester}  \& {Lyne}}{{Taylor}
  et~al.}{1993}]{Taylor93a}
{Taylor} J.~H.,  {Manchester} R.~N.,   {Lyne} A.~G.,  1993, \mn@doi [\apjs]
  {10.1086/191832}, \href {http://adsabs.harvard.edu/abs/1993ApJS...88..529T}
  {88, 529}

\bibitem[\protect\citeauthoryear{{Tomsick} \& {Muterspaugh}}{{Tomsick} \&
  {Muterspaugh}}{2010}]{Tomsick10}
{Tomsick} J.~A.,  {Muterspaugh} M.~W.,  2010, \mn@doi [\apj]
  {10.1088/0004-637X/719/1/958}, \href
  {http://adsabs.harvard.edu/abs/2010ApJ...719..958T} {719, 958}

\bibitem[\protect\citeauthoryear{{VERITAS} \& {MAGIC Collaborations}}{{VERITAS}
  \& {MAGIC Collaborations}}{2017}]{Veritas17}
{VERITAS} {MAGIC Collaborations} 2017, The Astronomer's Telegram, \href
  {http://adsabs.harvard.edu/abs/2017ATel10810....1V} {10810}

\bibitem[\protect\citeauthoryear{{Verbiest}, {Lorimer}  \&
  {McLaughlin}}{{Verbiest} et~al.}{2010}]{Verbiest10}
{Verbiest} J.~P.~W.,  {Lorimer} D.~R.,   {McLaughlin} M.~A.,  2010, \mn@doi
  [\mnras] {10.1111/j.1365-2966.2010.16488.x}, \href
  {http://adsabs.harvard.edu/abs/2010MNRAS.405..564V} {405, 564}

\bibitem[\protect\citeauthoryear{{Verbiest}, {Weisberg}, {Chael}, {Lee}  \&
  {Lorimer}}{{Verbiest} et~al.}{2012}]{Verbiest12}
{Verbiest} J.~P.~W.,  {Weisberg} J.~M.,  {Chael} A.~A.,  {Lee} K.~J.,
  {Lorimer} D.~R.,  2012, \mn@doi [\apj] {10.1088/0004-637X/755/1/39}, \href
  {http://adsabs.harvard.edu/abs/2012ApJ...755...39V} {755, 39}

\bibitem[\protect\citeauthoryear{{Wex}, {Johnston}, {Manchester}, {Lyne},
  {Stappers}  \& {Bailes}}{{Wex} et~al.}{1998}]{Wex98}
{Wex} N.,  {Johnston} S.,  {Manchester} R.~N.,  {Lyne} A.~G.,  {Stappers}
  B.~W.,   {Bailes} M.,  1998, \mn@doi [\mnras]
  {10.1046/j.1365-8711.1998.01700.x}, \href
  {http://adsabs.harvard.edu/abs/1998MNRAS.298..997W} {298, 997}

\makeatother
\end{thebibliography}

\bsp	
\label{lastpage}
\end{document}